\documentclass[twocolumn,showpacs,preprintnumbers,amsmath,amssymb]{revtex4}

\usepackage{amsmath,amssymb}

\usepackage{color}
\usepackage[dvips]{graphicx}
\usepackage{subfigure}

\def\gap{\;\rlap{\lower 2.5pt
 \hbox{$\sim$}}\raise 1.5pt\hbox{$>$}\;}
\def\lap{\;\rlap{\lower 2.5pt
   \hbox{$\sim$}}\raise 1.5pt\hbox{$<$}\;}
\def\gsim{\;\rlap{\lower 2.5pt
 \hbox{$\sim$}}\raise 1.5pt\hbox{$>$}\;}
\def\lsim{\;\rlap{\lower 2.5pt
   \hbox{$\sim$}}\raise 1.5pt\hbox{$<$}\;}
\def\msun{{M_\odot}}
\def\rsun{{R_\odot}}

\def\cm{{\rm\,cm}}
\def\Mpc{{\rm\,Mpc}}
\def\kpc{{\rm\,kpc}}

\def\GeV{{\rm\,GeV}}
\def\TeV{{\rm\,TeV}}
\def\sec{{\rm\,s}}
\def\sr{{\rm\,sr}}

\def\spose#1{\hbox to 0pt{#1\hss}}
\def\lta{\mathrel{\spose{\lower 3pt\hbox{$\mathchar''218$}}
     \raise 2.0pt\hbox{$\mathchar''13C$}}}
\def\gta{\mathrel{\spose{\lower 3pt\hbox{$\mathchar''218$}}
     \raise 2.0pt\hbox{$\mathchar''13E$}}}
\newcommand{\beq}{\begin{equation}}
\newcommand{\eeq}{\end{equation}}
\newcommand{\be}{\begin{equation}}
\newcommand{\ee}{\end{equation}}

\newcommand{\ls}{\mathrel{\raise1.16pt\hbox{$<$}\kern-7.0pt 
\lower3.06pt\hbox{{$\scriptstyle \sim$}}}}         
\newcommand{\gs}{\mathrel{\raise1.16pt\hbox{$>$}\kern-7.0pt 
\lower3.06pt\hbox{{$\scriptstyle \sim$}}}}         
\def\VEV#1{{\langle #1 \rangle}}
\long\def\comment#1{}

\def\msun{M_{\odot}}
\def\fun#1#2{\lower3.6pt\vbox{\baselineskip0pt\lineskip.9pt
  \ialign{$\mathsurround=0pt#1\hfil##\hfil$\crcr#2\crcr\sim\crcr}}}
\def\lap{\mathrel{\mathpalette\fun <}}
\def\gap{\mathrel{\mathpalette\fun >}}
\newcommand{\ba}{\begin{eqnarray}}
\newcommand{\ea}{\end{eqnarray}}

\begin{document}
\bibliographystyle{apsrev.bst}

\title{Neutralino annihilation into $\gamma$--rays in the Milky Way and in external galaxies} 
\thanks{Preprint numbers: DFTT 17/2004}

\author{N. Fornengo$^{1,2}$, L. Pieri$^{1}$, S. Scopel$^{1}$}
\email{fornengo@to.infn.it, pieri@to.infn.it, scopel@to.infn.it}
\affiliation{$^{1}$Dipartimento di Fisica Teorica, Universit\`a di Torino \\
via P. Giuria 1, I--10125 Torino, Italy} 
\affiliation{$^{2}$Istituto Nazionale di Fisica Nucleare, Sezione di Torino \\
via P. Giuria 1, I--10125 Torino, Italy} 

%
\begin{abstract}
\noindent
We discuss the gamma--ray signal from dark matter annihilation in our
Galaxy and in external objects, namely the Large Magellanic Cloud, the
Andromeda Galaxy (M31) and M87. We derive predictions for the fluxes
in a low energy realization of the Minimal Supersymmetric Standard
Model and compare them with current data from EGRET, CANGAROO-II and
HEGRA and with the capabilities of new--generation satellite--borne
experiments, like GLAST, and ground-based ${\rm \check{C}}$erenkov
telescopes, like VERITAS. We find fluxes below the level required to
explain the possible indications of a $\gamma$--ray excess shown by
CANGAROO-II (toward the Galactic Center) and HEGRA (from M87). As far
as future experiments are concerned, we show that only the signal from
the galactic center could be accessible to both satellite--borne
experiments and to ACTs, even though this requires very steep dark
matter density profiles.
\end{abstract}

\maketitle

\section{Introduction}
The nature of the Cold Dark Matter (CDM) which is believed to compose
galactic halos is probably the most important open issue in present
Cosmology. A popular solution to this puzzle is given by the lightest
supersymmetric particle (LSP) which, in most supersymmetry breaking
scenarios, is the neutralino $\chi$. In this case Dark Matter (DM)
would be not so dark after all, since $\chi$-$\chi$ annihilation is
expected to lead, among other final states, to a $\gamma$ signal which
could in principle be detected above known backgrounds.  In
particular, since the neutralino annihilation rate is proportional to
the square of its density, a signal enhancement is expected in high
density regions like the center of our Galaxy or that of
external ones, with the exciting possibility that such $\gamma$--rays
might be identified by forthcoming or just operating atmospheric
Cerenkov telescopes (ACT) such as VERITAS \cite{VERITAS}
HESS \cite{HESS} and MAGIC \cite{MAGIC} or by satellite-borne detectors
like GLAST \cite{GLAST}, let alone the even more intriguing chance
that a hint of an exotic source of $\gamma$--rays could actually be
already present in the data of existing experiments, like
EGRET \cite{egret_gc} or CANGAROO-II \cite{CANGAROO}.  However,
assessing the size of such signals depends on many uncertain aspects
of both astrophysics and particle physics. For instance, the central
structure of the DM halos is far from being well determined, and this
can lead to uncertainties in the calculation of expected $\gamma$
rates spanning several orders of magnitude.  Another sensitive issue
is the presence of substructures in galactic halos, which can change
predictions as compared to a smooth mass distribution.

The aim of the present paper is to investigate the possibility that
neutralino annihilations in the halo of our galaxy
\cite{lightindirect, gamma_papers,stoehr3,munnoz}, or that of external ones
\cite{Pieri:04,gamma_extra} (namely the Large Magellanic Cloud, the
Andromeda Galaxy and M87) could produce detectable fluxes of
$\gamma$--rays. To this purpose we will discuss present astrophysical
uncertainties and focus on deriving consistent predictions for these
fluxes in a specific realization of supersymmetry, the effective
Minimal Supersymmetric Standard Model (MSSM).

The plan of the paper is as follows: in Section II the main
ingredients for the calculation of the $\gamma$--ray flux from
neutralino annihilation are introduced; in Section III we discuss the
contribution to the flux calculation coming from astrophysics, while
in Section IV the contribution from particle physics is discussed, and
the effective MSSM Supersymmetric model is outlined. In Section V we
show our results and compare them to present data and the prospects of
future experiments; finally, Section VI is devoted to our conclusions.

\section{The $\gamma$-ray Flux}\label{sec:flux}

The diffuse photon flux from neutralino annihilation in the galactic
halo, coming from a given direction in the sky defined by the
angle--of--view $\psi$ from the Galactic Center, and observed by a
detector with angular resolution $\theta$ can be written as:
\begin{equation}
\frac{d \Phi_\gamma}{dE_\gamma}(E_\gamma, \psi, \theta) =
\frac{d \Phi^{\rm SUSY}} {dE_\gamma}(E_\gamma) \times \Phi^{\rm
cosmo}(\psi, \theta)
\label{flussodef}
\end{equation}
The energy dependence in Eq. (\ref{flussodef}) is given by the
annihilation spectrum:
\begin{equation}
\frac{d \Phi^{\rm SUSY}}{dE_\gamma}(E_\gamma) =  
  \frac{1}{4 \pi} \frac{\VEV{\sigma_{\rm ann} v}}{2 m^2_\chi} \cdot 
\sum_{f} \frac{d N^f_\gamma}{d E_\gamma} B_f  
\label{flussosusy}
\end{equation}
where $\VEV{\sigma_{\rm ann}v}$ is the neutralino self--annihilation
cross--section times the relative velocity of the two annihilating
particles, $d N^f_\gamma / dE_\gamma$ is the differential photon
spectrum for a given $f$-labeled annihilation final state with
branching ratio $B_f$ and $m_\chi$ denotes the neutralino mass. The
geometry--dependence is given by the line--of--sight integral, defined
as:
\begin{equation}
\Phi^{\rm cosmo}(\psi,\theta) = \int_{\Delta \Omega (\psi,\theta)}
d \Omega' \int_{\rm l.o.s} \rho_{\chi}^2 (r(\lambda,\psi')) d\lambda(r,\psi')
\label{flussocosmoMW}
\end{equation}
for the diffuse emission of our Galaxy, and
\begin{equation}
\Phi^{\rm cosmo}(\psi, \theta) =  
\frac{1}{d^2} 
\int_{0}^{\min[R_G,r_{\rm max}(\Delta \Omega)]}  4 \pi r^2 \rho_{\chi}^2(r) dr  
\label{flussocosmoextra}
\end{equation}
for the emission from an extragalactic object located at the direction
$\psi$. In Eq. (\ref{flussocosmoMW}), $\rho_\chi(r)$ is the dark
matter density profile, $r$ is the galactocentric distance, related to
the distance $\lambda$ from us by $r = \sqrt{\lambda^2 + \rsun^2 -2
\lambda \rsun \cos \psi}$ ($\rsun$ is the distance of the Sun from the
galactic center) and $\Delta \Omega (\psi,\theta)$ is the solid angle
of observation pointing in the direction of observation $\psi$ and for
an angular resolution of the detector $\theta$.  Moreover, in
Eq. (\ref{flussocosmoextra}) $d$ is the distance of the external
object from us, $R_G$ is the radius of the external galaxy and $r_{\rm
max}(\Delta \Omega)$ is the maximal distance from the center of the
external galaxy which is seen within the solid angle $\Delta \Omega
(\psi,\theta)$.

We focus our attention on the fact that Eq. (\ref{flussodef}) is
factorized into two distinct terms: a ``cosmological factor''
$\Phi^{\rm cosmo}$ which takes into account the geometrical
distribution of DM in the Universe, and a
``supersymmetric factor'' $\Phi^{\rm SUSY}$ which contains the
information about the nature of dark matter.  In Sections
\ref{sec:cosmofactor} and \ref{susyfactor} we will present results on
the two factors separately. 

\section{The ``Cosmological Factor''}\label{sec:cosmofactor}

In the following we present the determination of the ``cosmological
factor'' $\Phi^{\rm cosmo}$, as defined in Eqs. (\ref{flussocosmoMW})
and (\ref{flussocosmoextra}). The dependence of $\Phi^{\rm cosmo}$ on
the astrophysical and cosmological details that we explore here is
based on the determination of the shape of the dark matter halo. This
takes into account the possible existence and prominence of central
cusps, the study of the physical extent of the constant--density inner
core, and the possible presence of a population of sub--halos. We
remind the reader that, for the moment, no definitive answer can be
given to these questions by experimental constraints. In particular,
the discussion about the possible existence of a halo with a cuspy
behavior in its inner regions is still quite open. Moreover,
theoretical predictions differ substantially among themselves, or take
into account different input parameters.
 
These facts reflect themselves in a large uncertainty in the
predictions of the gamma--ray fluxes arising from $\Phi^{\rm cosmo}$,
as it is discussed and quantified in the following.

\subsection{Modeling the Dark Matter Halo} 

The modeling of the DM density profile is an open question. It can be
addressed through numerical N-body simulations whose scale resolution
is about few $\times 10^{-3} r_{100}$, where $r_{100}$ is defined as the
radius within which the halo average density is about $100 \rho_c$
($\rho_c$ is the critical density). The very inner slope of the
profile is then usually just extrapolated and does not take into
account interactions with the baryons which fall in the DM potential
well. A number of profiles have been proposed. Here we discuss some of
the profiles which are compatible with observations and which we will
use in our analysis.

In our calculation we mainly focus on the NFW profile (hereafter
NFW97) \cite{Navarro:97}
\begin{equation}
\rho_\chi^{\rm NFW97} = \frac{\rho^{\rm NFW97}_s}{\left (r/r^{\rm NFW97}_s \right )
\left (1 + r/r^{\rm NFW97}_s \right )^{2}}
\end{equation}
and the Moore et al. profile (M99) \cite{Moore:99}:
\begin{equation}
\rho_\chi^{\rm M99} = \frac{\rho_s^{\rm M99}}{\left (r/r_s^{\rm M99} 
\right )^{1.5} 
\left [1 + \left (r/r_s^{\rm M99}\right )^{1.5} \right ]}
\end{equation}
The scale radii $r^i_s$ and the scale densities $\rho^i_s$ ($i={\rm
NFW97}, {\rm M99}$) can be deduced by observations (the virial mass of
the halo or the rotation curves) and by theoretical considerations
that allow to determine the concentration parameter $c = r_{\rm
vir}/r_s$ (the virial radius $r_{\rm vir}$ is defined as the radius
within which the halo average density is $200 \rho_c$). The
concentration parameters, $c_{\rm NFW97}$ and $c_{\rm M99}=0.64 \
c_{\rm NFW97}$, have been computed according to Ref. \cite{ENS} with
the assumption of a CDM power spectrum with a shape parameter
$\Gamma=0.2$ normalized to $\sigma_8=0.9$.

In addition to the two profiles mentioned before, we include in our
predictions the conservative modified isothermal profile with a
constant density core (iso-core):
\begin{equation}
\rho_\chi^{\rm iso-core} = \frac{\rho_s^{\rm iso-core} }{
\left [1 + \left (r/r_s^{\rm iso-core}\right )^{2} \right ]}  
\end{equation}
and a profile which has been recently proposed by Moore and
collaborators (M04) \cite{Moore:04}:
\begin{equation}
\rho_\chi^{\rm M04} = \frac{\rho_s^{\rm M04}}{\left (r/r_s^{\rm M04} 
\right )^{1.16} \left (1 + r/r_s^{\rm M04} \right )^{1.84}}  
\end{equation}
\begin{figure}[t] 
\vspace{-20pt}
\includegraphics[height=8cm,width=8cm]{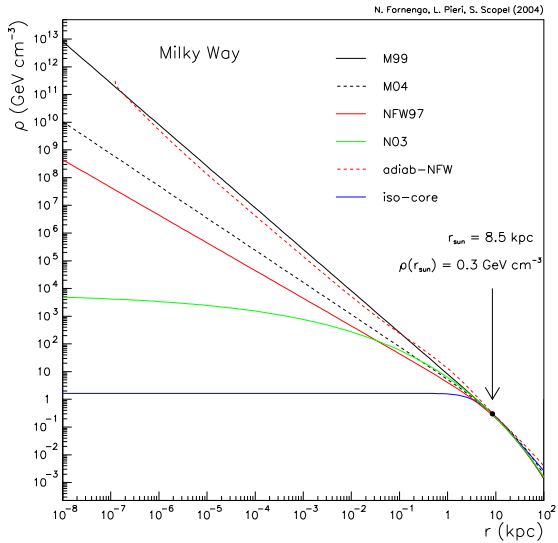}
\caption{Comparison between cuspy and cored dark matter density
profiles for the Milky Way, as a function of the distance from the
center of the Galaxy. All the curves are normalized to
$\rho_0\equiv\rho(R_\odot)=0.3$ GeV cm$^{-3}$. }
\label{fig2}
\end{figure}

Fig. \ref{fig2} shows the comparison among the above mentioned
profiles for the Milky Way, normalized to a local density of $0.3 \GeV
\cm^{-3}$ and to $\rsun = 8.5 \kpc$.  Two more profiles are shown for
comparison on Fig. \ref{fig2}. One is the numerical profile obtained
in Ref. \cite{ullio:01} when the adiabatic growth of a central black
hole is taken into consideration (adiab-NFW).  This hypothesis of
black-hole formation has been applied here to the NFW97 profile, and
the density profile has been normalized as previously mentioned. The
resulting profile has a behavior at the galactic center which is
similar to the one of the M99 profile, therefore we won't discuss it
in more details. The last profile which is shown in the figure is a
cored one recently obtained in Ref. \cite{Navarro:03} (N03):
\begin{equation}
\rho_\chi^{\rm N03} = \rho_s^{\rm N03} \exp\left [ -\frac{2}{\alpha} \left [ \left (
 \frac{r}{r_s^{N03}} \right )^{\alpha} - 1 \right] \right]
\end{equation}
where $\alpha =0.17$, $r_s^{\rm N03} = r^{\rm NFW97}_s$ and
$\rho_s^{\rm N03}= \rho^{\rm NFW97}_s/4$. As noticed in
Ref. \cite{Moore:04}, this profile is compatible with the M04 as far
as the resolution of the N-body simulation holds. In the inner part of
the Galaxy, it is an extrapolation which postulates the existence of a
constant density core. Another recently proposed profile which does
not exhibit singular behaviour, and which has been shown to be able to
reproduce to a good precision the rotational velocities of low surface
brightness galaxies \cite{stoehr2}, is given in
Ref. \cite{stoehr1}. Predictions of gamma--ray fluxes for this profile
are given in Ref.  \cite{stoehr3}.

Since profiles shallower than the NFW97 hardly give observable
fluxes of photons, we will not discuss it in detail. Studying the
cored halos, we will limit ourselves to the iso-core profile, which is
pretty conservative.

\begin{table}[t]
\begin{center}
\begin{tabular}{|c|c|c|c|}
\hline
~ Galaxy ~ & ~~ mass ($\msun$)~~  & ~~ distance (Kpc)~~  & ~  $r_{\rm vir}$ (Kpc) ~ \\
\hline 
MW & $1.0 \cdot 10^{12}$ & 8.5 & 205  \\ 
LMC & $1.4 \cdot 10^{10}$ & 49 & 49  \\ 
M31 & $2.0 \cdot 10^{12}$  & 770 &  258 \\ 
\hline
\end{tabular}
\caption{\label{tab1} Masses, distances 
and virial radii for the Milky Way, the LMC and M31.}
\end{center}
\end{table}
\begin{table}[t]
\begin{center}
\begin{tabular}{|c|c|c|}
\hline
Profile & scale radius $r_s$ (Kpc) & scale density $\rho_s$ ($\msun \  \kpc^{-3}$) \\
\hline
NFW97 & 21.746 & $5.376 \cdot 10^{6}$ \\
M99 & 34.52 & $1.060 \cdot 10^{6}$ \\
M04 & 32.625 & $2.541 \cdot 10^{6}$ \\
iso-core & 4 & $7.898 \cdot 10^{6}$ \\
\hline
\end{tabular}
\caption{\label{tabMW} Scale radii and scale densities for the NFW97,
M99, M04 and iso-core density profiles calculated for the Milky Way.}
\end{center}
\end{table}
\begin{table}[t]
\begin{center}
\begin{tabular}{|c|c|c|}
\hline
Profile & scale radius $r_s$ (Kpc) & scale density $\rho_s$ ($\msun \  \kpc^{-3}$) \\
\hline
NFW97 & 4.353 & $8.50 \cdot 10^{6}$ \\
M99 & 6.8 & $1.80 \cdot 10^{6}$ \\
M04 & 6.426 & $3.22 \cdot 10^{6}$ \\
iso-core & 1.5 & $2.17 \cdot 10^{7}$ \\
\hline
\end{tabular}
\caption{\label{tabLMC} Scale radii and scale densities for the
NFW97, M99, M04 and iso--core density profiles calculated for the LMC.}
\end{center}
\end{table}
\begin{table}[t]
\begin{center}
\begin{tabular}{|c|c|c|}
\hline
Profile & scale radius $r_s$ (Kpc) & scale density $\rho_s$ ($\msun \  \kpc^{-3}$) \\
\hline
NFW97 & 30.271 & $4.20 \cdot 10^{6}$ \\
M99 & 47.298 & $0.86 \cdot 10^{6}$ \\
M04 & 44.697 & $1.55 \cdot 10^{6}$ \\
iso-core & 4 & $7.898 \cdot 10^{6}$ \\
\hline
\end{tabular}
\caption{\label{tabM31} Scale radii and scale densities for the
NFW97, M99, M04 and iso--core density profiles calculated for M31.}
\end{center}
\end{table}

Integrating the squared density along the line of sight introduces
divergences when cuspy profiles are considered. Therefore we enforce a
cut--off radius $r_{\rm cut}$ to the density profile, with a constant
density core therein. The smallest value for the cut--off radius which
we will use is $r_{\rm cut} =10^{-8} \kpc$, a value we will discuss in
the next Section, where the effect of varying $r_{\rm cut}$, both for
our Galaxy and for the external ones, will be discussed.

The analysis of Ref. \cite{Pieri:04} shows that a number of external
galaxies shine above the Galactic foreground. In the following we will
focus on the two most prominent galaxies at large angles with respect
to the Galactic Center, namely the Large Magellanic Cloud (LMC) and
the Andromeda Galaxy (M31) \cite{Pieri:04}.  Table \ref{tab1} shows
the astrophysical parameters for the Milky Way, the LMC and M31, while
Tables \ref{tabMW}, \ref{tabLMC} and \ref{tabM31} show the scale
radius ans the scale density parameters used in our calculations.

\subsubsection{Comment on the experimental constraints on the inner part of galaxies} 

As we have seen, theoretical estimates of the inner slope $\alpha$ of
the DM density profile $\rho(r) \propto r^{-\alpha}$ are still
uncertain. Moreover, observations which should constrain the $\alpha$
parameter do not give clear and definitive answers on its value.  A
number of works give in fact non-unique values for the slope.

In Ref.  \cite{buote:03} spatially resolved spectra of the diffuse hot
(X-rays) gas of galaxies and clusters measured with the Chandra
satellite were used to infer the radial mass distribution of the
considered systems.  An analysis was done on 2 clusters which are
relaxed in their cores on ${\it O} (10^2 \kpc)$ to ${\it O} (\Mpc)$
scales and do not have strong radio sources in their center. Resulting
values for $\alpha$ are 1.25 and 1.35. A value of $\alpha$ less than
1 is found when disturbed X-ray surface brightness clusters are
used. Yet the X-ray method uses the double assumption of a single
phase gas in hydrostatic equilibrium, which for instance is
questionable in the central regions where rapid cooling occurs.

Other studies of radial mass profiles inferred by the radial profile
of the intracluster medium density and temperature measured with
Chandra can be found in Ref. \cite{bautz:03} where the analysis of 5
clusters gives $1 < \alpha < 2$.

Different results are found by Ref. \cite{deBlok:03} using Low Surface
Brightness (LSB) Galaxies rotation curves. Fits to their measured
curves give a mean value $\VEV{\alpha} = 0.2$, although tails in the
distribution extend further, up to $\alpha = 2$. In
Ref. \cite{sand:03} a combination of strong--lensing data and
spectroscopic measurements of stellar dynamics of the brightest
cluster galaxies was used to derive values of $\alpha$. Three
clusters, containing both radial and tangential arcs, have been found.
The obtained distribution gives $\VEV{\alpha} = 0.52$ with
$\Delta\alpha = 0.3$.

In Ref. \cite{Hayashi:03} the full radial extent of LSB galaxies
rotation curves, instead of its inner portion, was used to determine
the inner slope of the DM density profile. Convergence criteria for
the N-body simulations taken from Ref. \cite{Power:03} give a minimum
radius for which simulations are reliable $r_{\rm conv} = 1 h^{-1}
\kpc$. It is shown that, at that radius, $2/3$ of the sample in
Ref. \cite{deBlok:03} is consistent with a profile which lies between
the simulated NFW97 and M99 ones. There are inconsistencies with CDM
predictions in those galaxies which show a sharp transition between
the rising and flat part of the rotation curve. This is due to the
fact that rotation curves of gas disks are compared with the
spherically-averaged circular velocity profiles of DM halos. This
assumption may not be correct in non regular galaxies.

Another study of high resolution $H_\alpha$ rotation curves for dwarf
and LSB galaxies has been recently carried out in
Ref. \cite{swaters:03}. In that work it is shown that rotation curves
data are insufficient to rule out halos with $\alpha = 1$, although
none of the galaxies require an inner cuspy profile instead of a core
density feature. Results on $\alpha$ range from 0 to 1.2, although the
quality of the fit is good only up to $\alpha = 1$. Other analysis on
large sets of data of high--resolution rotation curves also show
consistency with cored mass distributions \cite{isorefs}.

An indirect estimate of $\alpha$ can be inferred through the weak
gravitational lensing measurements of X-ray luminous
clusters \cite{dahle:03}: one finds $0.9 < \alpha < 1.6$.

The analysis of the microlensing optical depth toward the Galactic
Center was performed in Ref. \cite{merrifield:03}. Assuming a na\"ive
spherically symmetric profile normalized to our position in the Milky
Way, the authors find $\alpha = 0.4$. They argue that the value
$\alpha = 1$ can be reached by considering a flattened halo with a
ratio of polar to equatorial axis of 0.7.

\subsection{Including the Effect of the Inner Core} \label{sec:core}

There exists a physical minimal radius, $r_{\rm cut}$, within which the
self--annihilation rate $t_{l} \sim (\VEV{\sigma_{\rm ann} v} \ n_\chi
(r_{\rm cut}))^{-1}$ equals the dynamical time $t_{\rm dyn} \sim (G
\bar{\rho} )^{- \frac{1}{2}}$ \cite{Berezinsky:92}, where $\bar{\rho}$
is the mean halo density and $n_\chi$ is the neutralino number
density. When this procedure is applied to the density profiles we are
using, the evaluated $r_{\rm cut}$ are of the order of
$10^{-8}-10^{-9} \kpc$ for the M99 profile and of $10^{-13}-10^{-14}
\kpc$ for a NFW97. Evaluating the constant core is indeed a much more
complicate issue. Taking into account additional effects, especially
tidal interactions, the central core of galaxies can significantly exceed
the values quoted above, reaching values as large as ${\cal O}(0.1
- 1) \kpc$ \cite{Berezinsky:03}. We want to remind that also numerical
simulations, from which the cuspy behavior is deduced for the inner
parts by means of extrapolation, are actually testing the halo shape
down to ${\cal O}(0.1)\kpc$ \cite{Moore:04,Navarro:03}.

In our analysis we will take into account this large uncertainty in
the inner core radius by varying $r_{\rm cut}$ in the range $[10^{-8},
10^{-1}] \kpc$.

\subsection{Results for $\Phi^{\rm cosmo}$}

\begin{figure}[t] 
\vspace{-20pt}
\includegraphics[height=8cm,width=8cm]{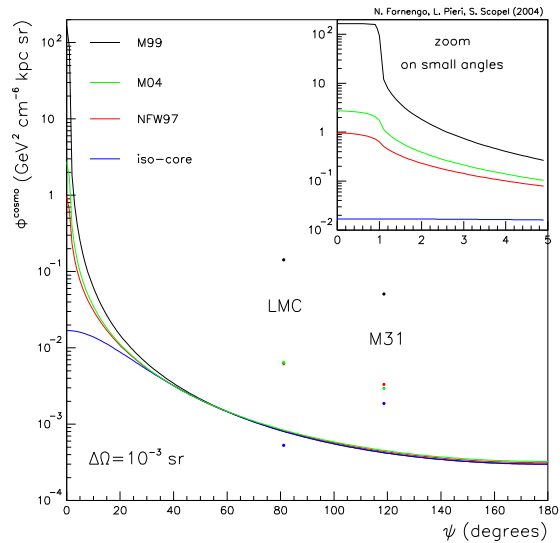}
\includegraphics[height=8cm,width=8cm]{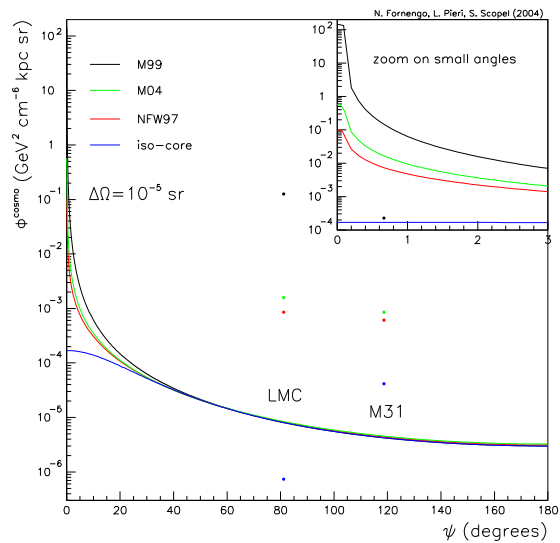}
\caption{The lines denote the ``cosmological factor'' $\Phi^{\rm
cosmo}$ for the Milky Way, calculated for different dark matter
profiles, for a solid angle $\Delta \Omega = 10^{-3}$ sr (upper panel)
and $\Delta \Omega = 10^{-5}$ sr (lower panel). The small boxes show a
zoom at small angles toward the galactic center. A constant--density
central region of radius $r_{\rm cut}=10^{-8} \kpc$ has been used for
the cuspy profiles. The points at $\psi \simeq 81^\circ$ and $\psi
\simeq 119^\circ$ denote the values of $\Phi^{\rm cosmo}$ for LMC and
M31, respectively. From top to bottom the points refer to different
halo profiles: Moore, NFW97, M04, iso-core in the upper panel; Moore,
M04, NFW97, iso-core in the lower panel.}
\label{fig2a}
\end{figure}

The results of the calculations of the cosmological factor $\Phi^{\rm
cosmo}$ for the Milky Way are shown in Fig. \ref{fig2a}, for the four
main profiles previously discussed and for a detector with angular
resolution equal to $1^\circ$ and $0.1^\circ$. A constant--density
central region of radius $r_{\rm cut}=10^{-8} \kpc$ has been used for
the cuspy profiles. Since the value of $r_{\rm cut}$ used in
Fig.\ref{fig2a} somehow represents a lower bound on the acceptable
values of this parameter, the values of $\Phi^{\rm cosmo}$ shown in
Fig. \ref{fig2a} can be taken as an upper bound on the cosmological
factor, for any given halo profile and for the two representative
acceptance angles. Clearly the non--cuspy profiles are not affected by
the choice of $r_{\rm cut}$.

In the same figure, the values of $\Phi^{\rm cosmo}$ for LMC and M31
are also shown. We see that these external galaxies can be resolved
against the galactic signal in all cases, except for the case of LMC
with an iso-core density profile. These two external galaxies can
therefore be looked at as gamma--ray sources from DM annihilation
(provided that the ensuing gamma--ray flux can be detected against the
gamma--ray background). If a gamma--ray signal were detected, for
instance from the galactic center, it should be correlated to a
corresponding signal both from LMC and from M31. Since the
``supersymmetric factor'' is the same for all the sources, the
relative strength of the gamma--ray fluxes from the galactic center,
LMC and M31, could then be used to deduce information on the halo
shape, since it depends only on the DM density profile. However, this
possibility is strongly limited by the fact that $\Phi^{\rm cosmo}$
for LMC and M31 is much smaller than the one from the galactic center,
as is clear from Fig. \ref{fig2a}. The ensuing fluxes from external
galaxies will therefore be much smaller than the ones from the
galactic center.

\begin{figure}[t] 
\vspace{-20pt}
\includegraphics[height=8cm,width=8cm]{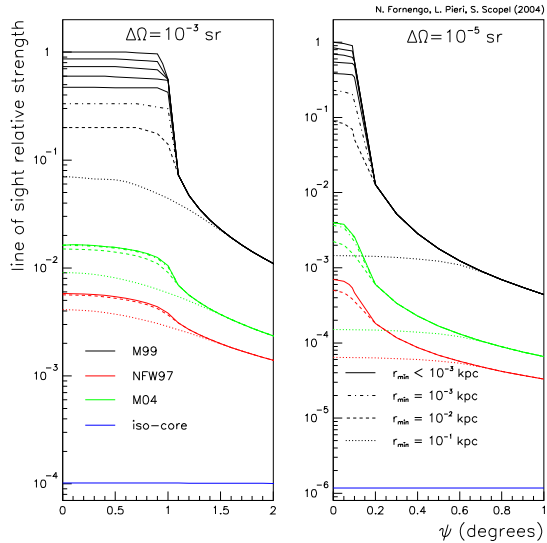}
\caption{Relative strength of the line--of--sight integral with
respect to different halo profiles and different inner core radii for
the Milky Way. Numbers are normalized to the highest value of the
$\Phi^{\rm cosmo}$ given by a M99 profile with a physical cut--off
radius of $10^{-8}$ kpc and at $\psi=0$. Left panel: solid
angle $\Delta \Omega = 10^{-3}$ sr. Right panel: solid angle $\Delta
\Omega = 10^{-5}$ sr.}
\label{fig3}
\end{figure}
\begin{figure}[t] 
\vspace{-20pt}
\includegraphics[height=8cm,width=8cm]{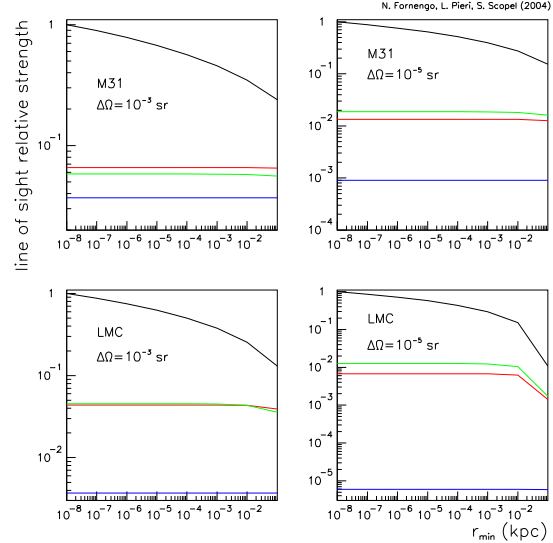}
\caption{Relative strength of the line--of--sight integral with
respect to different halo profiles and different inner core radii for
M31 (upper panels) and the LMC (lower panels). Numbers are normalized
to the highest value of the $\Phi^{\rm cosmo}$ given by a M99 profile
with a physical cut-off radius of $10^{-8}$ kpc and at $\psi=0$. Left
panels: solid angle $\Delta \Omega = 10^{-3}$ sr. Right panels: solid
angle $\Delta \Omega = 10^{-5}$ sr.}
\label{fig3a}
\end{figure}

The dependence of the cosmological factor on the cut--off radius of
the inner core is shown in Fig. \ref{fig3} for the Milky Way and in
Fig. \ref{fig3a} for LMC and M31. In these figures we plot the ratio
$\Phi^{\rm cosmo} ({\rm profile}, r_{\rm cut},\psi)/\Phi^{\rm cosmo}
({\rm M99}, r_{\rm cut}= 10^{-8} \kpc, \psi=0)$ for the M99, NFW97,
M04 and iso-core profiles and for $r_{\rm cut}$ in the range discussed
above 
. As expected, a cored distribution or a less
cuspy profile than the M99 decreases the cosmological factor by a
significant amount with respect to the most optimistic hypotheses of a
M99 profile with an inner core radius $r_{\rm cut}= 10^{-8}
\kpc$. Figs. \ref{fig3} and \ref{fig3a} quantify this effect.

In the case of the Milky Way, the reduction factor at the galactic
center can be sizeable: for instance, when a NFW97 profile with $r_{\rm
cut}=0.1$ kpc is used, the reduction is of the order of $4 \cdot
10^{-3}$ for a solid angle of observation $\Delta \Omega = 10^{-3}$ sr
and $6 \cdot 10^{-5}$ for $\Delta \Omega = 10^{-5}$ sr. In the case of
the iso--cored distribution the reduction factor is as large as
$10^{-4}$ for $\Delta \Omega = 10^{-3}$ and $10^{-6}$ for $\Delta
\Omega = 10^{-5}$.

The same trend is observed for the external galaxies which we have
considered, although the net effect is less prominent. In the case of
M31, the reduction is at most a few $\times 10^{-2}$ for $\Delta
\Omega = 10^{-3}$ and it can reach $10^{-3}$ for $\Delta \Omega =
10^{-5}$ and the iso--core profile. For LMC, the reduction is again of
the order of $10^{-2}$--$10^{-3}$, except for the iso--core profile
and $\Delta \Omega = 10^{-5}$, for which it reaches values
of the order of $10^{-5}$.

In the following, for definiteness we will refer to the most
optimistic values of $\Phi^{\rm cosmo}$ shown in Fig. \ref{fig2a},
obtained for a M99 profile with a cut--off radius of $10^{-8} \kpc$
and to a NFW97 shape, with the same cut--off radius. Results for
different halo profiles or core parameters can be easily obtained by
scaling the results according to Figs. \ref{fig3} and \ref{fig3a}.

\subsection{Including Substructures} 

In the CDM scenario, sub-halos that accrete into larger systems are
tidally stripped of a fraction of their mass, originating debris
streams \cite{helmi:03}. Their dense central cores, however, survive
the merging event and continue to orbit within the parent halo.  High
resolution N-body simulations \cite{Moore:99,Ghigna:99} have indeed
shown that DM halos host a population of sub-halos with a distribution
function depending on the sub-halo mass and on the distance of the
sub--halo from the halo center \cite{Blasi:00}. 

The effect of including sub--halos in the Milky Way and in the
galaxies of the Local Group has been discussed in Refs.
\cite{aloisio:02,Roldan:00,Pieri:04,PiBr:04}, where different
parameters for the sub-halo distribution, along with the existence of
mass stripping and tidal heating, have been considered, and a minimum
mass of $10^{6} \msun$ was assumed for the sub--halos.  The existence
of such a sub--halo population leads to average boost factors for
expected rates which depend on the modeling of the sub--halo
distribution and on the density profile, and can range from few
unities to more than $10^{4}$. In no case, however, the field of view
toward the Galactic Center is affected, since in that region the
gravitational strengthening reduces the probability of finding
sub--halos.  The total effect of the presence of sub-halos in external
galaxies is limited to a factor 2--5.

A discussion on the minimum mass of sub--halos in our Galaxy can be
found in Ref. \cite{Berezinsky:03}, where small scale clumps are
considered, with masses down to $10^{-8} \msun$ for a DM constituted
by neutralinos.  An average enhancement factor of 2--5 is found,
depending on the profile, while the enhancement toward the Galactic
Center is found to be of a factor $\sim0.3$ (NFW97) to $\sim0.5$ (M99).

Hereafter we consider values for the ``cosmological factor'' related
to an unclumpy scenario. For a clumpy halo our results can be scaled
according to the previous considerations.

\section{The ``Supersymmetric Factor''}\label{susyfactor}

In our study we employ the MSSM supersymmetric extension of the
Standard Model, which is defined as an effective theory at the
electroweak scale. The scheme is defined in terms of a minimal number
of parameters, only the ones which are necessary to shape the
essential properties of the theoretical structure of the MSSM and of
its particle content. A number of assumptions are therefore imposed at
the electroweak scale: a) all squark soft--mass parameters are
degenerate: $m_{\tilde q_i} \equiv m_{\tilde q}$; b) all slepton
soft--mass parameters are degenerate: $m_{\tilde l_i} \equiv m_{\tilde
l}$; c) all trilinear parameters vanish except those of the third
family, which are defined in terms of a common dimensionless parameter
$A$: $A_{\tilde b} = A_{\tilde t} \equiv A m_{\tilde q}$ and
$A_{\tilde \tau} \equiv A m_{\tilde l}$. In addition, we employ also
the standard relation at the electroweak scale between the $U(1)$ and
$SU(2)$ gaugino mass parameters: $M_1 = (5/3) \tan^2\theta_W \simeq
0.5~M_2$, which is the low energy consequence of an underlying
unification condition for the gaugino masses at the GUT scale. In this
class of gaugino--universal models, the neutralino mass has a lower
bound of about 50 GeV. This limit is induced by the lower bound on the
chargino mass determined at LEP2 \cite{susylep2}: $m_{\chi^\pm} \gsim
100$ GeV. This is at variance with respect to effective MSSM schemes
which do not posses gaugino-universality, where the neutralino mass
can be as low as few GeV's (see for instance
Refs. \cite{light,lightindirect} and references quoted
therein). Gamma--ray detection from the annihilation of these light
neutralinos has also been analized in Ref. \cite{munnoz}, in the
context of SUGRA models where gaugino non universality is defined at
the GUT scale.

Due to the above mentioned assumptions, the supersymmetric parameter
space of our scheme consists of the following independent parameters:
$M_2$, $\mu$, $\tan\beta$, $m_A$, $m_{\tilde q}$, $m_{\tilde l}$ and
$A$. In the previous list of parameters $\mu$ denotes the Higgs mixing
mass parameter, $m_A$ is the mass of the CP-odd neutral Higgs boson
and $\tan\beta\equiv v_t/v_b$ is the ratio of the two Higgs v.e.v.'s
that give mass to the top and bottom quarks.

When we perform a numerical random scanning of the supersymmetric
parameter space, we employ the following ranges for the parameters: $1
\leq \tan \beta \leq 50$, $100~ {\rm GeV }\leq |\mu|$, $M_2 \leq
6000~{\rm GeV }$, $100~{\rm GeV}\leq m_{\tilde q}, m_{\tilde l} \leq
3000~{\rm GeV }$, ${\rm sign}(\mu)=-1,1$, $90~{\rm GeV }\leq m_A \leq
1000~{\rm GeV }$, $-3 \leq A \leq 3$. The range on both $M_2$ and
$\mu$ extends up to 6 TeV in order to allow us to study also very
heavy neutralinos, with a mass up to about 3 TeV.

The parameters space of our effective MSSM is constrained by many
experimental bounds: accelerators data on supersymmetric and Higgs
boson searches \cite{higgslep2} and on the invisible width of the $Z$
boson, measurements of the branching ratio of the $b\rightarrow s +
\gamma$ decay and of the upper bound on the branching ratio of $B_s
\rightarrow \mu^+ + \mu^-$, measurements of the muon anomalous
magnetic moment $a_\mu \equiv (g_{\mu} - 2)/2$.  The limits we use
are: $2.18 \cdot 10^{-4} \leq BR (b \rightarrow s + \gamma) \leq 4.28
\cdot 10^{-4}$ \cite{bsgamma_exp}; $BR(B_s \rightarrow \mu^+ + \mu^-)
< 7.5 \cdot 10^{-7}$ (95\% C.L.) \cite{bsmumu_exp}; $-142 \leq \Delta
a_{\mu} \cdot 10^{11} \leq 474 $ (this 2$\sigma$ C.L. interval takes
into account the recent evaluations of Refs. \cite{davier,hagiwara2}).

For the theoretical evaluation of
$BR (b \rightarrow s + \gamma)$ and $BR(B_s \rightarrow \mu^+ +
\mu^-)$ we have used the results of Ref. \cite{bsgamma_theory} and
Ref. \cite{bsmumu_theory}, respectively, with inclusion of the QCD
radiative corrections to the bottom--quark Yukawa coupling discussed
in Ref. \cite{carena}.  We notice that gluinos do not enter directly
into our loop contributions to $BR(b \rightarrow s + \gamma)$ and
$BR(B_s \rightarrow \mu^+ + \mu^-)$, since we assume flavor-diagonal
sfermion mass matrices. Gluinos appear only in the QCD radiative
corrections to the $b$ Yukawa coupling: in this case $M_3$ is taken at
the standard unification value $M_3 = M_2 \;
\alpha_3(M_Z)/\alpha_2(M_Z)$, where $\alpha_3(M_Z)$ and
$\alpha_2(M_Z)$ are the SU(3) and SU(2) coupling constants evaluated
at the scale $M_Z$. 

Another relevant observational constraint comes from Cosmology. The
recent observations on the cosmic microwave background from
WMAP \cite{wmap}, used in combination with galaxy surveys,
Lyman--$\alpha$ forest data and the Sloan Digital Sky Survey
Collaboration results \cite{sloan}, are leading to a precise knowledge
of the cosmological parameters, and in particular of the amount of
dark matter in the Universe. From the analysis of Ref. \cite{wmap}, we
can derive a restricted range for the relic density of a cold species
like the neutralinos. The density parameter of cold dark matter is
bounded at $2\sigma$ level by the values: $(\Omega_{CDM} h^2)_{\rm
min} = 0.095$ and $(\Omega_{CDM} h^2)_{\rm max} = 0.131$. This is the
range for CDM that we consider in the present paper. For
supersymmetric models which provide values of the neutralino relic
abundance $\Omega_{\chi} h^2$ smaller than the minimal value
$(\Omega_{CDM} h^2)_{\rm min}$, i.e. for models where the neutralino
represents a subdominant DM component, we accordingly rescale
the value of the DM density: $\rho_\chi(r) = \xi \rho(r)$ with $\xi
=\Omega_{\chi} h^2/(\Omega_{CDM} h^2)_{\rm min}$.

We recall that the relic abundance $\Omega_{\chi} h^2$ is essentially
given by $\Omega_{\chi} h^2 \propto \langle\sigma_{\rm ann}
v\rangle^{-1}_{\rm int}$, where $\langle\sigma_{\rm ann} v\rangle_{\rm
int}$ is the thermal--average of the product of the neutralino
annihilation cross section and velocity, integrated from the
freeze--out temperature in the early Universe down to the present
one. The analytical calculation of $\sigma_{\rm ann}$ relies on the
full set of available final states: fermion-antifermion pairs, gluon
pairs, Higgs boson pairs, one Higgs boson and one gauge boson, pairs
of gauge bosons \cite{anni}. We have not included coannihilation
\cite{coannihilation} in our evaluation of the neutralino relic
abundance, since in our effective supersymmetric model a matching of
the neutralino mass with other particle masses is usually accidental,
and not induced by some intrinsic relationship among the different
parameters of the supersymmetric model, like instead in a constrained
SUGRA scheme, The inclusion of coannihilation would not change the
main results of our analysis, since it would only reflect in a limited
reshuffle of a small fraction of the points of the scatter plots
displayed in the next Sections.

\subsection{The annihilation cross section}

\begin{figure}[t] 
\vspace{-20pt}
\includegraphics[height=8cm,width=8cm]{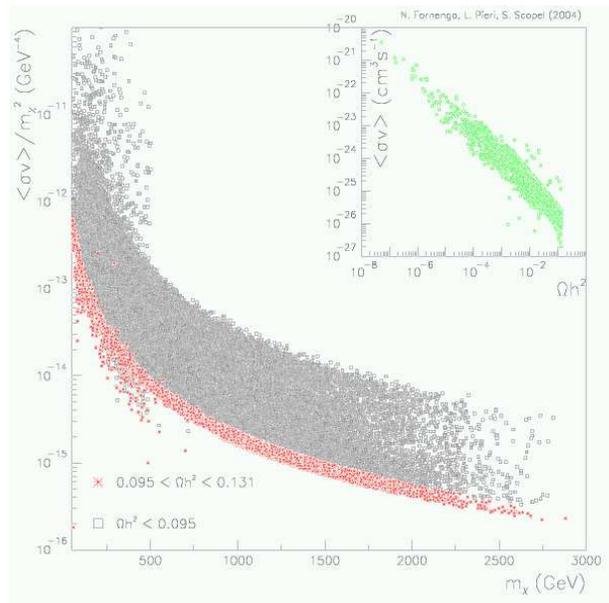}
\caption{The thermally--averaged annihilation cross--section divided by
the square of the neutralino mass $m_\chi$ as a function of $m_\chi$
in the frame of the eMSSM. Crosses show the WMAP-preferred zone for a
DM dominant neutralino. In the small box the annihilation cross
section at the present epoch is shown as a function of the neutralino
relic abundance.}
\label{fig4}
\end{figure}

As already stated in Sec. \ref{sec:flux}, the gamma--ray flux produced
by neutralino annihilation depends on the thermal average of the
neutralino self--annihilation cross section $\langle\sigma_{\rm ann}
v\rangle$ in the galactic halo at present time. The behaviour of
$\langle\sigma_{\rm ann} v\rangle/m_\chi^2$, which is a relevant
quantity in the calculation of the gamma--ray flux, is shown in
Fig. \ref{fig4} as a function of the neutralino mass and for the
effective MSSM we are using. We remind that $\langle\sigma_{\rm ann}
v\rangle$ in general is different from $\langle\sigma_{\rm ann}
v\rangle_{\rm int}$ which is responsible for the determination of
the relic abundance. The two cross sections closely follow each other
only for s--wave annihilation. An inverse proportionality between the
gamma--ray signal and the relic abundance is therefore usually a good
approximation, although deviations are present. This effect is shown
in the box--insert in Fig. \ref{fig4}.

Other key ingredients for the determination of the gamma--ray signal
are the branching ratios of the annihilation cross section into the
different final states. For neutralino lighter than 1 TeV the
branching ratios were shown in Ref. \cite{pbar_susy}. Fig. \ref{fig4}
extends the behaviour of the branching ratios for neutralino masses
higher that one TeV. We see that in this case the dominant channels
are the two gauge bosons and the gauge boson+Higgs boson final states.

\begin{figure}[t] 
\vspace{-20pt}
\includegraphics[height=8cm,width=8cm]{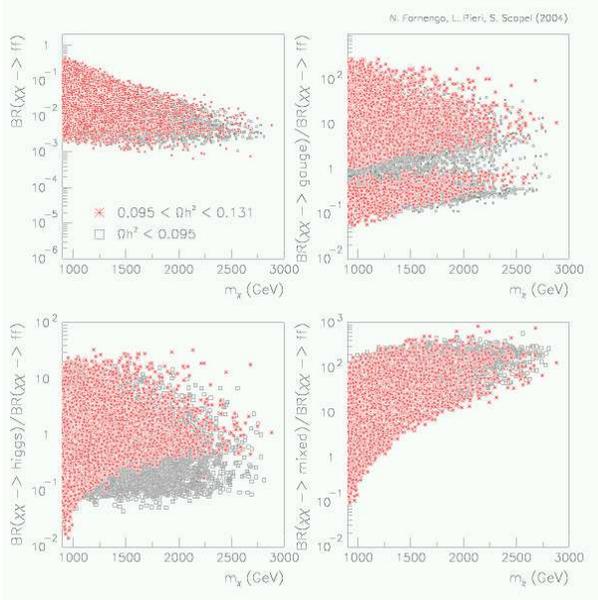}
\caption{
Branching ratios for high mass neutralino annihilation into 
fermions (upper left panel) and relative strength of annihilation into 
gauge bosons (upper right), Higgs bosons (lower left) and a gauge boson
and a Higgs boson (lower right) with respect to the annihilation into fermions.
}
\label{fig4a}
\end{figure}

\subsection{The Photon Spectrum}

The diffuse photon spectrum from neutralino annihilation originates
from the production of fermions, gauge bosons, Higgs bosons and
gluons. Both gauge bosons and Higgs bosons eventually decay into
fermions. The hadronization of quarks and gluons, in addition to
radiative processes, can produce $\gamma$--rays. The main channel of
production of $\gamma$--rays goes through the production and subsequent
decay of neutral pions. The contribution to the $\gamma$--ray spectrum
from production and decay of mesons other than pions (mostly $\eta$,
$\eta^{\prime}$, charmed and bottom mesons) and of baryons is usually
subdominant as compared to $\pi^0$ decay and it has been neglected.
Neutralino annihilation into lepton pairs can also produce $\gamma$--rays
from electromagnetic showering of the final state leptons. This
process can be dominant for $E_{\gamma}\lsim$ 100 MeV, when the
neutralino annihilation process has a sizable branching ratio into
lepton pairs. In the case of production of $\tau$ leptons, their
semihadronic decays also produce neutral pions, which then further
contribute to the gamma--ray flux.

As discussed in Ref. \cite{lightindirect}, we have evaluated the
gamma--ray fluxes originating from hadronization and radiative
processes by means of a Monte Carlo simulation with the PYTHIA package
\cite{pythia}. In the present paper we extend that analysis by giving
explicit fits to our numerical distributions which are valid for
the energies of interest in the current analysis, i.e. for photon
energies $E> 10$ GeV. When a flux is presented for energies below 10
GeV, the numerical analysis has been used.

The differential spectra of photons from DM annihilation have been
parametrized as follows:
\beq \frac{dN^{i}_\gamma}{dx} = \eta
x^{a} {\rm e}^{b+cx+dx^2+ex^3}
\label{dndxqg}
\eeq 
where $x = E_\gamma / m_\chi$ and $i$ identify quarks, $W$, $Z$ and
gluons. The value of $\eta$ is 2 for $W$, $Z$ and top quark final
states, and 1 otherwise. In the case of $\tau$ leptons, the functional
form for the differential number of photons is:
\beq
\frac{dN^{\tau}_\gamma}{dx} = x^{a_\tau} (b_\tau x+c_\tau x^2+d_\tau
x^3) {\rm e}^{e_\tau x}
\label{dndxtau}
\eeq

The values of the parameters of the fits are given in Tables
\ref{paraqg} and \ref{paratau} for the two representative values of
$m_\chi = 500 \GeV$ and $m_\chi = 1 \TeV$. 

\begin{figure}[t] 
\vspace{-20pt}
\includegraphics[height=8cm,width=8cm]{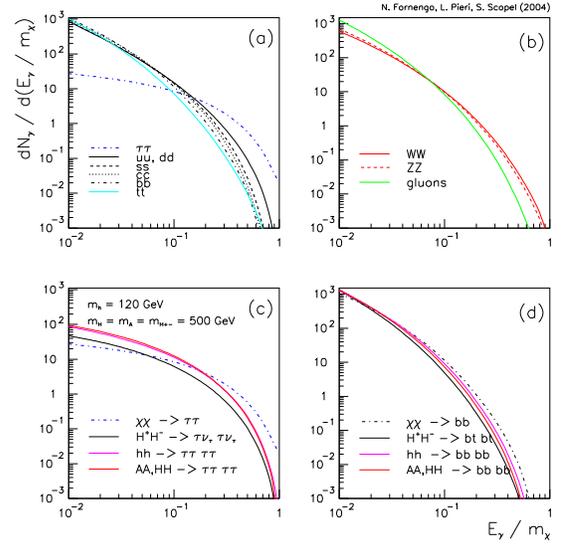}
\caption{The photon spectrum from a $m_\chi=1$ TeV neutralino
annihilation into: (a) leptons, (b) gauge bosons, (c) Higgs bosons
decaying into $\tau$'s and (d) Higgs bosons decaying into $b$'s. For
each curve a branching ratio of 100\% in that channel has been
considered.}
\label{fig5}
\end{figure}

In Fig. \ref{fig5} we show an example of photon spectra originated by
neutralino annihilation into different pure final states of a
neutralino with $m_\chi=1$ TeV. We see that at lower energies the
dominant contribution is given by the $\gamma$--rays coming from the
hadronization of quarks and gluons. The spectra coming from gauge
bosons are somewhat harder, while the hardest ones are given by the
$\tau$ lepton. In the case of Higgs bosons, the spectra are mainly
driven by the type of particle in which the Higgs bosons decays, and
are somewhat softer.

\begin{table}[t]
\vskip 1cm
\begin{center}
\begin{tabular}{|c||c|c|c||c|c|c|}
\hline
& \multicolumn{3}{|c||}{$m_\chi = 500 \GeV$} 
& \multicolumn{3}{|c|}{$m_\chi = 1 \TeV$} \\ \hline \hline 
& $u$ & $s$& $t$ & $u$ & $s$& $t$ \\ \hline 
a & -1.5 & -1.5 & -1.5 & -1.5 & -1.5 & -1.5 \\ 
b & 0.047 & 0.093 & -0.44 & 0.0063 &  0.040 &  -0.45 \\
c & -8.70 & -9.13 & -19.50 & -8.62 & -8.84 & -19.05 \\
d & 9.14 & 4.49 & 22.96 & 8.53 & 2.77 & 21.96 \\
e & -10.30 & -9.83 & -16.20 & -9.73 & -7.71 & -15.18 \\ \hline
& $d$ & $c$& $b$ & $d$ & $c$& $b$ \\ \hline 
a & -1.5 & -1.5 & -1.5 & -1.5 & -1.5 & -1.5 \\ 
b & 0.047 & 0.25 & 0.48 & 0.0063 &  0.17 &  0.37 \\
c & -8.70 & -10.76 & -16.87 & -8.62 & -10.23 & -16.05 \\
d & 9.14 & 4.25 & 21.09 & 8.53 & 2.13 & 18.01 \\
e & -10.30 & -8.70 & -22.49 & -9.73 & -7.00 & -19.50 \\ \hline
& $W$ & $Z$& $g$ & $W$ & $Z$& $g$ \\ \hline
a & -1.5 & -1.5 & -1.5 & -1.5 & -1.5 & -1.5 \\
b & -0.85 & -0.76 & 0.55 & -0.95 &  -0.83 &  0.48 \\ 
c & -11.07 & -11.96 & -20.78 & -9.86 & -11.175  & -20.51  \\
d & 9.47 & 8.65 & 26.79 & 6.25 & 6.5902 & 24.42 \\
e & -6.80 & -5.21 & -22.80 & -4.37 & -3.6468 & -19.56 \\
\hline
\end{tabular}
\caption{\label{paraqg} Fitted parameters of Eq. (\ref{dndxqg}) for
the annihilation of neutralinos into quarks and gauge bosons,
calculated for $m_\chi = 500 \GeV$ and $m_\chi = 1 \TeV$. Fits
obtained with these parameters are valid down to $E = 10$ GeV.}
\end{center}
\end{table}

\begin{table}[t]
\vskip 1cm
\begin{center}
\begin{tabular}{|c|c|c|}
\hline
& $m_\chi = 500 \GeV$ & $m_\chi = 1 \TeV$ \\ \hline 
$a_\tau$ & -1.34 & -1.31  \\ 
$b_\tau$ &  6.27 &  6.94  \\
$c_\tau$ &  0.89 & -4.93  \\
$d_\tau$ &  -4.90 & -0.51 \\
$e_\tau$ &  -5.10 & -4.53  \\ 
\hline
\end{tabular}
\caption{\label{paratau} 
Parameters of Eq. (\ref{dndxtau}) for the annihilation of neutralinos
into $\tau$ leptons, calculated for $m_\chi = 500 \GeV$ and 
$m_\chi = 1 \TeV$. Fits obtained with these parameters are valid
down to $E = 10$ GeV.}
\end{center}
\end{table}

\subsection{Resuls on $\Phi^{\rm SUSY}$}

\begin{figure}[t] 
\vspace{-20pt}
\includegraphics[height=8cm,width=8cm]{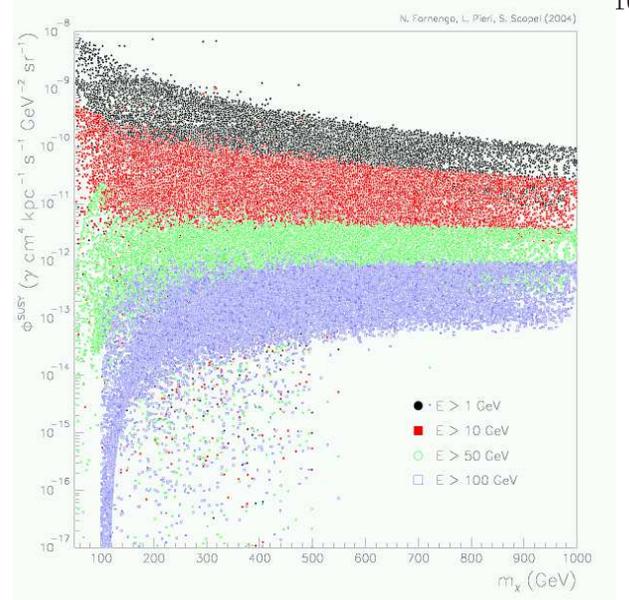}
\caption{ The ``supersymmetric factor'' $\Phi^{\rm SUSY}$ as a
function of the neutralino mass. Different colors show different
threshold energies above which the energy spectra have been integrated.}
\label{fig6}
\end{figure}

The composition of the information on the neutralino annihilation
cross--section and branching ratios with the informations coming from
the differential spectra of photons from the annihilation of
neutralinos in pure final states provides the prediction of what we
have called the ``supersymmetric factor'' $\Phi^{\rm SUSY}$ in the
$\gamma$-ray flux computation.  Fig. \ref{fig6} shows $\Phi^{\rm
SUSY}$ defined as the integral of the gamma--ray flux of
Eq. (\ref{flussosusy}) above a set of sample threshold energies: 1,
10, 50 and 100 GeV. When the threshold energy is small, $\Phi^{\rm
SUSY}$ is rougly inversely proportional to the neutralino mass. Since
the neutralinos annihilate almost at rest, when the threshold energy
increases $\Phi^{\rm SUSY}$ significantly drops because the highest
available photon energy is $E \sim m_\chi$ for any given neutralino
mass. The highest value for $\Phi^{\rm SUSY}$ is of the order of
$10^{-8}$ cm$^4$ kpc$^{-1}$ s$^{-1}$ GeV$^{-2}$ sr$^{-1}$ when the
threshold energy is 1 GeV. At masses larger than about 500 GeV for any
given threshold energy the values of $\Phi^{\rm SUSY}$ all lie inside
a band which span no more than one order of magnitude. This makes the
predictions on the gamma--ray fluxes for large neutralino masses quite
predictive: the possible variation due to the different supersymmetric
models is confined to a relatively restricted range, much smaller than
for the case of lighter neutralinos.

The information on the factor $\Phi^{\rm SUSY}$ is detailed in Tables
\ref{ngamma1a}, \ref{ngamma1b} and \ref{ngamma2} where we give the
number of photons produced in each pure final state for different
threshold energies. This information may be used to make predictions
for the gamma--ray fluxes also for DM candidates other than the
neutralino.

The results of this Section and of Sec. \ref{sec:cosmofactor} will be
used in the next Sections to predict the photon fluxes from the
galactic center and from our representative external galaxies.

\begin{table}[t]
\vskip 1cm
\begin{center}
\begin{tabular}{|c|c|c|c|}
\hline
& $m_\chi = 500 \GeV$ & $m_\chi = 800 \GeV$ & $m_\chi = 1 \TeV$ \\ \hline 
\hline
\multicolumn{4}{|c|}{$\chi \chi \longrightarrow u \bar u \ (d \bar d)$} \\ \hline 
$N_\gamma (> 10 \GeV)$ & 6.65 & 9.79 & 11.63 \\
$N_\gamma (> 50 \GeV)$ & 0.91 & 1.78 & 2.37 \\
$N_\gamma (> 100 \GeV)$ & 0.23 & 0.59 & 0.87 \\
$N_\gamma (> 500 \GeV)$ & 0.00 & 1.9 $\times 10^{-3}$ & 8.4 $\times 10^{-3}$ \\ \hline 
\multicolumn{4}{|c|}{$\chi \chi \longrightarrow s \bar s$} \\ \hline 
$N_\gamma (> 10 \GeV)$ & 6.61 & 9.83 & 11.71 \\
$N_\gamma (> 50 \GeV)$ & 0.76 & 1.62 & 2.21 \\
$N_\gamma (> 100 \GeV)$ & 0.15 & 0.46 & 0.73 \\
$N_\gamma (> 500 \GeV)$ & 0.00 & 2.1 $\times 10^{-4}$ & 1.7 $\times 10^{-3}$ \\ \hline
\multicolumn{4}{|c|}{$\chi \chi \longrightarrow c \bar c$} \\ \hline 
$N_\gamma (> 10 \GeV)$ & 7.10 & 10.61 & 12.71 \\
$N_\gamma (> 50 \GeV)$ & 0.69 & 1.60 & 2.19 \\
$N_\gamma (> 100 \GeV)$ & 0.11 & 0.41 & 0.66 \\
$N_\gamma (> 500 \GeV)$ & 0.00 & 8.7 $\times 10^{-5}$ & 8.4 $\times 10^{-4}$ \\ \hline
\multicolumn{4}{|c|}{$\chi \chi \longrightarrow t \bar t$} \\ \hline 
$N_\gamma (> 10 \GeV)$ & 5.03 & 8.65 & 10.81 \\
$N_\gamma (> 50 \GeV)$ & 0.29 & 0.84 & 1.29 \\
$N_\gamma (> 100 \GeV)$ & 0.04 & 0.17 & 0.30 \\
$N_\gamma (> 500 \GeV)$ & 0.00 & 1.7 $\times 10^{-4}$ & 8.1 $\times 10^{-4}$ \\ \hline
\multicolumn{4}{|c|}{$\chi \chi \longrightarrow b \bar b$} \\ \hline 
$N_\gamma (> 10 \GeV)$ & 7.02 & 11.02 & 13.31 \\
$N_\gamma (> 50 \GeV)$ & 0.49 & 1.26 & 1.83 \\
$N_\gamma (> 100 \GeV)$ & 0.07 & 0.28 & 0.47 \\
$N_\gamma (> 500 \GeV)$ & 0.00 & 5.8 $\times 10^{-5}$ & 6.3 $\times 10^{-4}$ \\ \hline 
\multicolumn{4}{|c|}{$\chi \chi \longrightarrow {\rm gluons}$} \\ \hline 
$N_\gamma (> 10 \GeV)$ & 6.42 & 10.69 & 13.18 \\
$N_\gamma (> 50 \GeV)$ & 0.34 & 0.95 & 1.47 \\
$N_\gamma (> 100 \GeV)$ & 0.04 & 0.17 & 0.32 \\
$N_\gamma (> 500 \GeV)$ & 0.00 & 5.7 $\times 10^{-5}$ & 4.3 $\times 10^{-4}$ \\ 
\hline
\end{tabular}
\caption{\label{ngamma1a} Integrated number of photons above a given
energy $E$ from the annihilation of neutralinos with masses 500 GeV,
800 GeV and 1 TeV, for different channels of annihilation.}
\end{center}
\end{table}

\begin{table}[t]
\vskip 1cm
\begin{center}
\begin{tabular}{|c|c|c|c|}
\hline
& $m_\chi = 500 \GeV$ & $m_\chi = 800 \GeV$ & $m_\chi = 1 \TeV$ \\ \hline 
\hline
$N_\gamma (> 500 \GeV)$ & 0.00 & 5.8 $\times 10^{-5}$ & 6.3 $\times 10^{-4}$ \\ \hline 
\multicolumn{4}{|c|}{$\chi \chi \longrightarrow \tau^+ \tau^-$} \\ \hline 
$N_\gamma (> 10 \GeV)$ & 2.19 & 2.38 & 2.46 \\
$N_\gamma (> 50 \GeV)$ & 1.16 & 1.55 & 1.72 \\
$N_\gamma (> 100 \GeV)$ & 0.58 & 0.98 & 1.28 \\
$N_\gamma (> 500 \GeV)$ & 0.00 & 3.3 $\times 10^{-2}$ & 8.2 $\times 10^{-2}$ \\ \hline  
\multicolumn{4}{|c|}{$\chi \chi \longrightarrow W^+ W^-$} \\ \hline 
$N_\gamma (> 10 \GeV)$ & 4.76 & 7.15 & 8.45 \\
$N_\gamma (> 50 \GeV)$ & 0.52 & 1.14 & 1.57 \\
$N_\gamma (> 100 \GeV)$ & 0.11 & 0.34 & 0.52 \\
$N_\gamma (> 500 \GeV)$ & 0.00 & 1.3 $\times 10^{-3}$ & 4.4 $\times 10^{-3}$ \\ \hline 
\multicolumn{4}{|c|}{$\chi \chi \longrightarrow ZZ$} \\ \hline 
$N_\gamma (> 10 \GeV)$ & 4.96 & 7.67 & 9.19 \\
$N_\gamma (> 50 \GeV)$ & 0.48 & 1.12 & 1.57 \\
$N_\gamma (> 100 \GeV)$ & 0.09 & 0.30 & 0.49 \\
$N_\gamma (> 500 \GeV)$ & 0.00 & 0.9 $\times 10^{-3}$ & 3.1 $\times 10^{-3}$ \\ \hline 
\hline
\end{tabular}
\caption{\label{ngamma1b} Integrated number of photons above a given
energy $E$ from the annihilation of neutralinos with masses 500 GeV,
800 GeV and 1 TeV, for different channels of annihilation.}
\end{center}
\end{table}

\begin{table}[t]
\vskip 1cm
\begin{center}
\begin{tabular}{|c|c|c|}
\hline
\multicolumn{3}{|c|}{$\chi \chi \longrightarrow$ Higgs} \\ \hline \hline
& $h h \rightarrow b \bar b$ &
$h h \rightarrow \tau^+ \tau^-$ 
 \\ \hline 
$N_\gamma (> 10 \GeV)$ & 13.95 & 3.89 \\
$N_\gamma (> 50 \GeV)$ & 1.56 & 1.97 \\
$N_\gamma (> 100 \GeV)$ & 0.34 & 1.07  \\
$N_\gamma (> 500 \GeV)$ & 1.8 $\times 10^{-4}$ & 0.02  \\ \hline 
& $A A (H H) \rightarrow b \bar b$ &
$A A (H H) \rightarrow \tau^+ \tau^-$ 
 \\ \hline 
$N_\gamma (> 10 \GeV)$ & 13.32 & 4.00 \\
$N_\gamma (> 50 \GeV)$ & 1.37 & 2.01  \\
$N_\gamma (> 100 \GeV)$ & 0.27 & 1.09 \\
$N_\gamma (> 500 \GeV)$ & 6.8 $\times 10^{-5}$  & 0.02 \\ \hline 
& $H^+ H^- \rightarrow b \bar b$ &
$H^+ H^- \rightarrow \tau^+ \tau^-$ 
 \\ \hline
$N_\gamma (> 10 \GeV)$ & 11.41 & 2.00 \\
$N_\gamma (> 50 \GeV)$ & 1.00 & 1.00  \\
$N_\gamma (> 100 \GeV)$ & 0.17 & 0.54 \\
$N_\gamma (> 500 \GeV)$ & 7.4 $\times 10^{-5}$  & 0.01 \\ 
\hline 
\end{tabular}
\caption{\label{ngamma2} Integrated number of photons above a given
energy $E$ from the annihilation of 1 TeV neutralino into a sample
state of Higgs bosons, with subsequent decay into $b$ quarks or tau
leptons. A mass of 120 GeV has been assumed for the light Higgs, while
a mass of 500 GeV has been taken for the charged, heavy and
pseudoscalar Higgses.}
\end{center}
\end{table}

\section{Prediction and Detectability of Photon Fluxes}\label{flux}

In this Section we will show the results on the prediction of 
photon fluxes from neutralino annihilation in our Galaxy and in some
selected external galaxies. We will therefore study the detectability
of such signals with ground-based ${\rm \check{C}}$erenkov telescopes
and next generation satellite-borne experiments.

\subsection{Predicted Photon Fluxes from Neutralino Annihilation}

\begin{figure}[t] 
\vspace{-20pt}
\includegraphics[height=8cm,width=8cm]{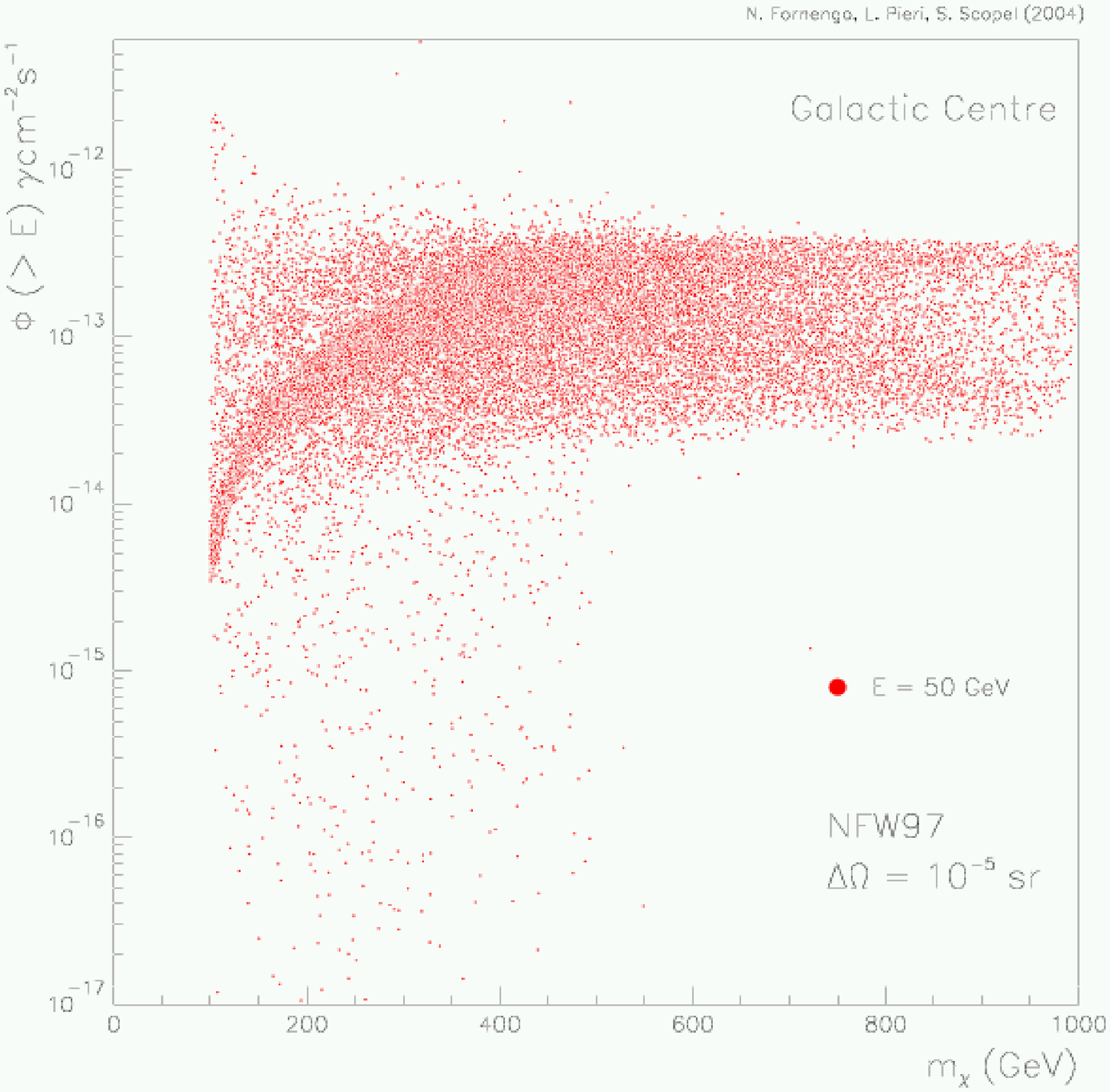}
\includegraphics[height=8cm,width=8cm]{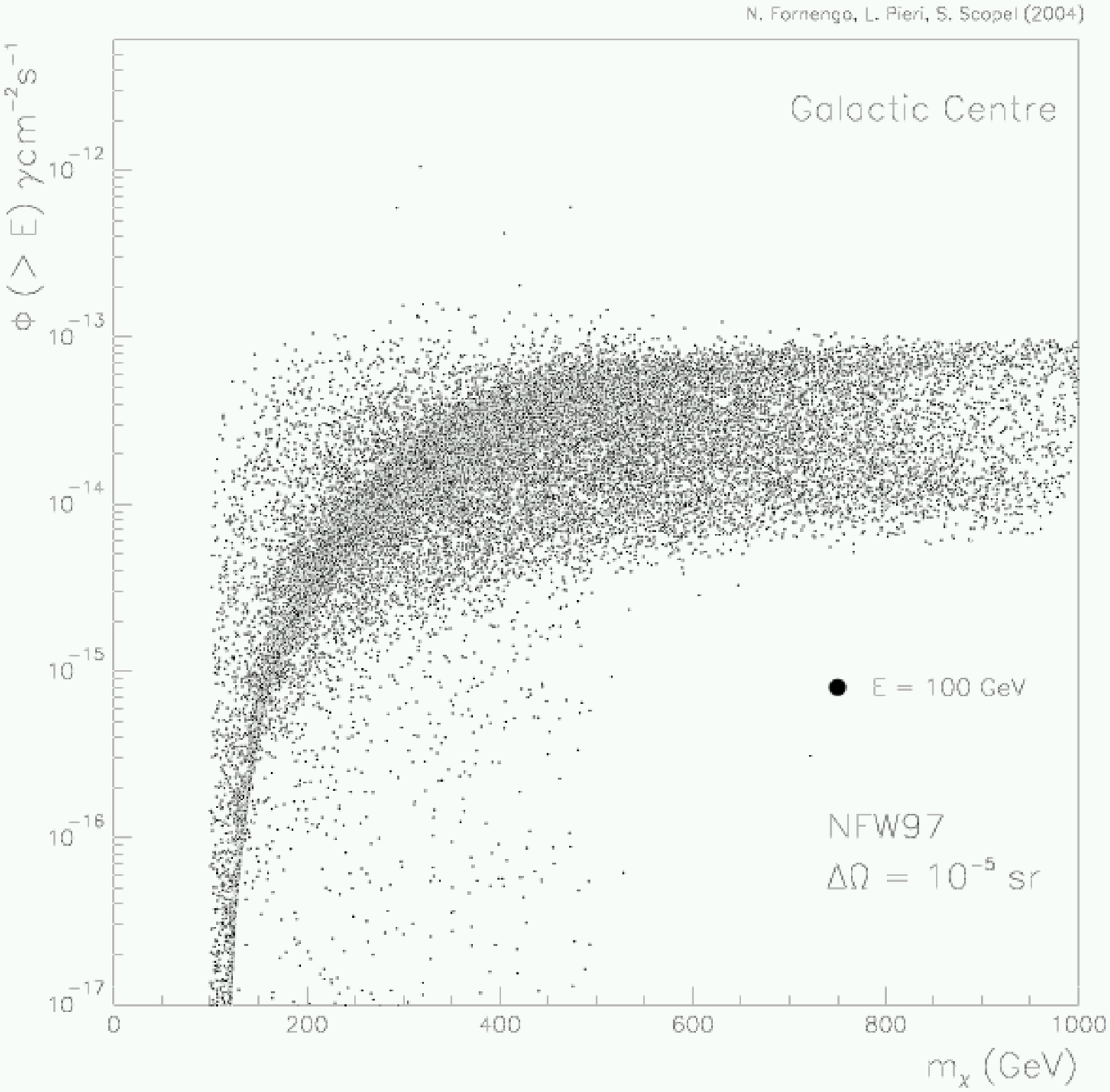}
\caption{Integrated gamma--ray fluxes from neutralino annihilation at
the galactic center, for a NFW97 density profile and inside a solid
angle $\Delta \Omega = 10^{-5}$ sr. Two representative threshold
energies have been assumed: 50 GeV (upper panel) and 100 GeV (lower
panel).}
\label{fig7}
\end{figure}

\begin{figure}[t] 
\vspace{-20pt}
\includegraphics[height=8cm,width=8cm]{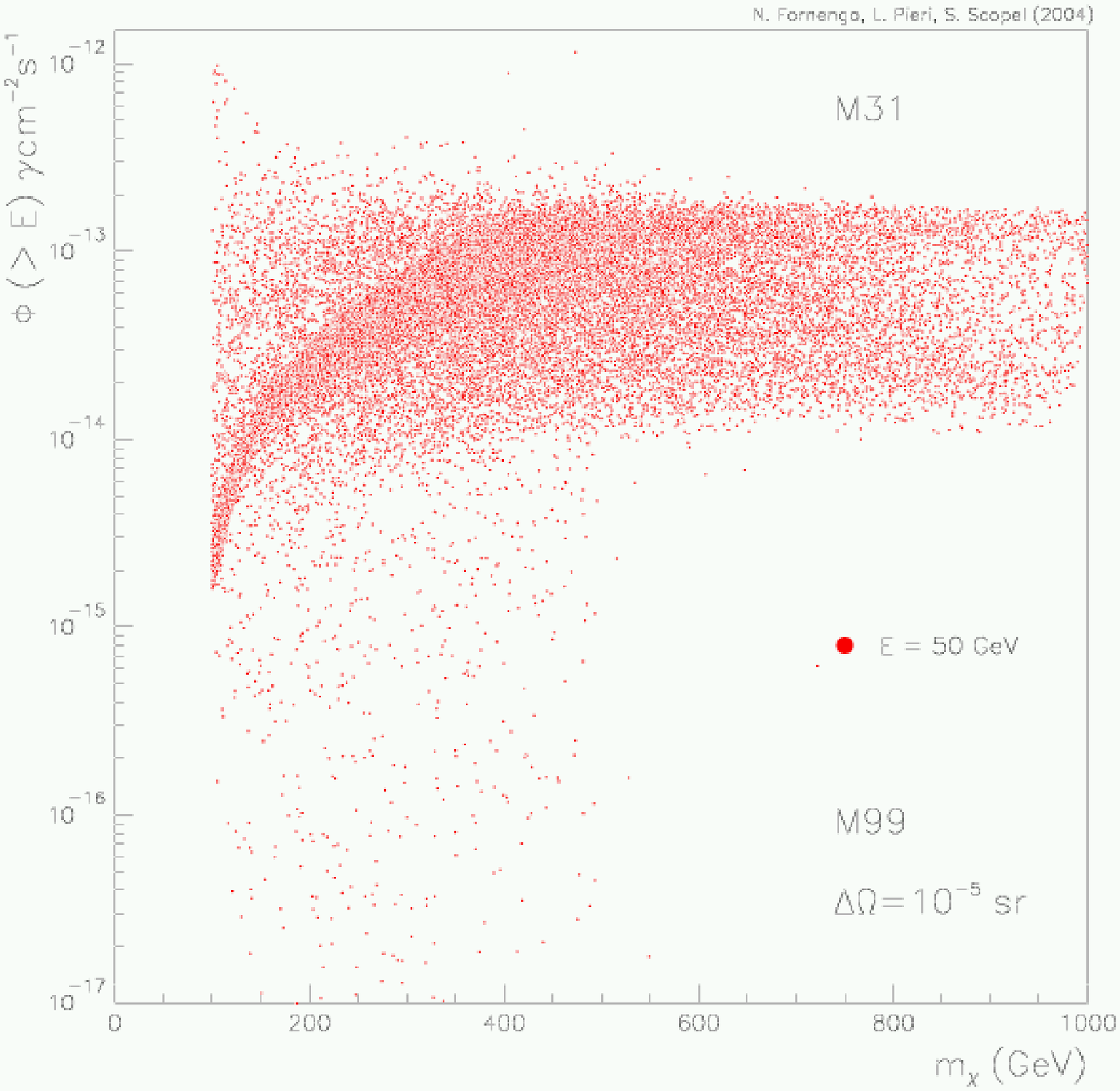}
\includegraphics[height=8cm,width=8cm]{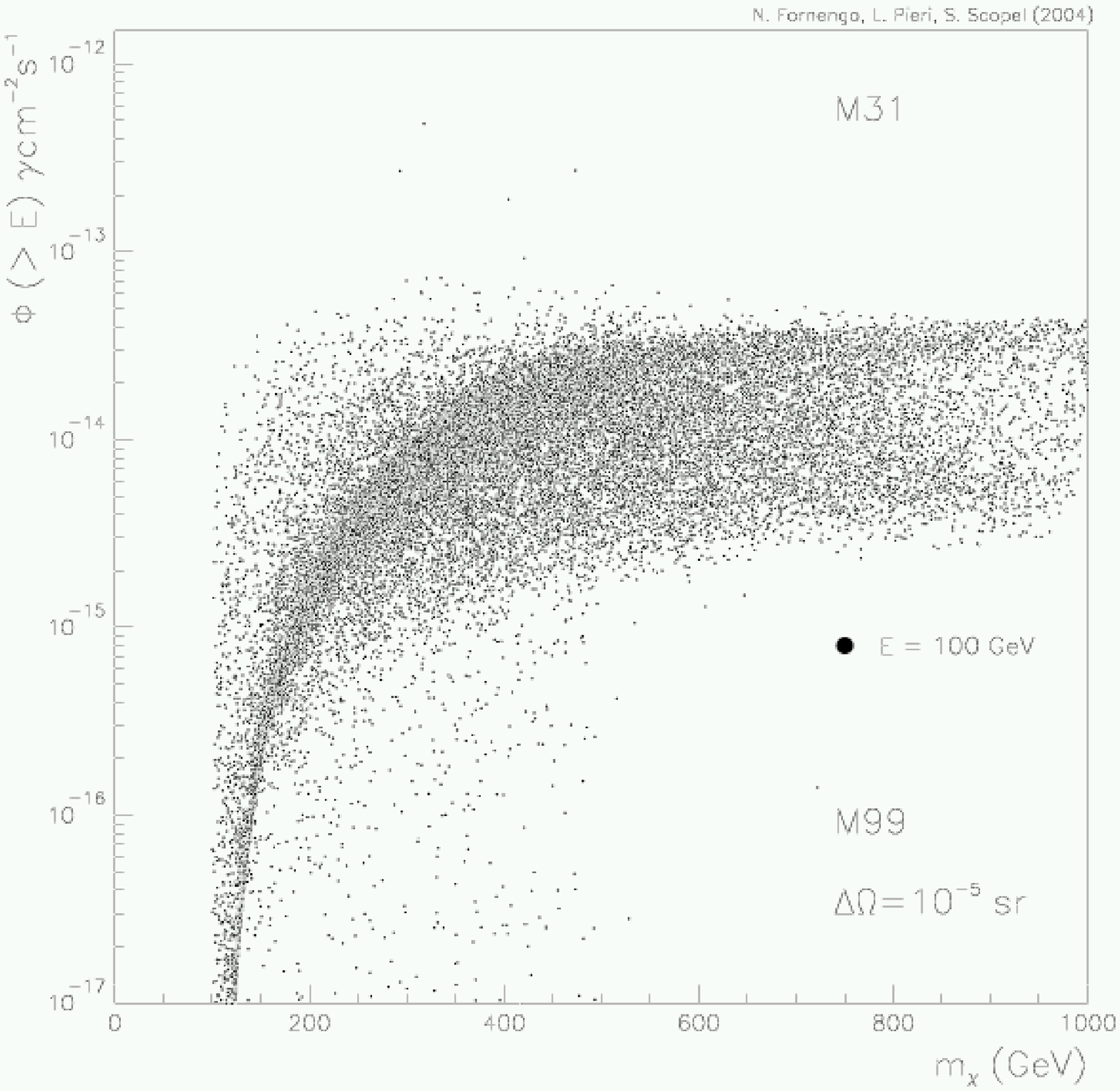}
\caption{Integrated gamma--ray fluxes from neutralino annihilation in
M31, for a M99 density profile and inside a solid angle $\Delta \Omega
= 10^{-5}$ sr. Two representative threshold energies have been
assumed: 50 GeV (upper panel) and 100 GeV (lower panel).}
\label{fig7a}
\end{figure}

In the previous Sections we have computed the ``cosmological factor''
$\Phi^{\rm cosmo}$ (see Fig. \ref{fig2a}) and the ``supersymmetric
factor'' $\Phi^{\rm SUSY}$ (see Fig. \ref{fig6}). We are now ready to
predict the gamma--ray fluxes fron neutralino annihilation in the
effective MSSM. Results are reported in Figs. \ref{fig7} and
\ref{fig7a}, where we show the expected fluxes of $\gamma$--rays with
energies above 50 GeV and 100 GeV from the galactic center and M31 and
for a detector aperture of $\Delta \Omega = 10^{-5}$ sr.
Fig. \ref{fig7} refers to the galactic center for a Milky Way with a
NFW97 density profile, while Fig. \ref{fig7a} is calculated for M31
with a M99 density profile. The spread of points is given by the
different SUSY parameters corresponding to each point.

In the case of the flux from the galactic center with a NFW97 profile
and a typical threshold energy of 50 GeV, we predict a maximal
gamma--ray flux of the order of $10^{-12}$ cm$^{-2}$ s$^{-1}$ for
neutralinos lighter than 200 GeV, while heavier neutralinos can
provide a maximal flux of the order of a few $10^{-13}$ cm$^{-2}$
s$^{-1}$. In the case of a M99 density profile toward the galactic
center, the fluxes are increased by a factor of about 160, as can be
deduced from Fig. \ref{fig3}. In this case the maximal fluxes can
reach the level of $10^{-10}$ cm$^{-2}$ s$^{-1}$. If the detector
threshold energy is increased to 100 GeV the gamma--ray fluxes are one
order of magnitude smaller. Finally, as a consequence of the
previously discussed property of $\Phi^{\rm SUSY}$, we see that for
neutralino masses heavier than about 500 GeV the supersymmetric models
we are considering provide gamma--ray fluxes inside a band with a
lower limit of a few $10^{-14}$ cm$^{-2}$ s$^{-1}$, for a NFW97
profile.  Obviously, if we enlarge the allowed intervals for the MSSM
parameters (our definitions are given in Sec. \ref{susyfactor}), lower
gamma--ray fluxes can be obtained also for heavy neutralinos. However,
if we consider natural mass scales for the supersymmetric model, which
means that we should not increase the scale of the mass parameters of
the model much over the TeV scale, Fig. \ref{fig7} shows the level of
the lower limit on the gamma--ray flux for heavy neutralinos.

Also the Andromeda Galaxy can provide gamma--ray fluxes of the order
of $10^{-12}$--$10^{-13}$ cm$^{-2}$ s$^{-1}$ inside a solid angle of
$\Delta \Omega = 10^{-5}$ sr, but only for a M99 density
profile. These values therefore represent the maximal fluxes which can
be produced by neutralino annihilation in M31. We remind that although
the galactic center is much brighter for the same density profile,
M31 can be resolved over the galactic gamma--ray signal due to its
location at $\psi=119^\circ$, as is shown in Fig. \ref{fig2a}.

In the following we will compare our expected fluxes with the
sensitivity curves of foreseeable experiments.

\subsection{Detectability of Photon Fluxes from Neutralino Annihilation}

We have considered two platforms of observations of $\gamma$--rays from
neutralino annihilation, corresponding to a ${\rm \check{C}}$erenkov
apparatus with the characteristics of VERITAS \cite{VERITAS} and to a
satellite--borne experiment similar to GLAST \cite{GLAST}.  The
detectability of the diffuse flux from DM annihilation is computed by
comparing the number $n_\gamma$ of expected $\gamma$ events with the
fluctuations of background events $n_{\rm bkg}$. To this purpose we
define the following ratio $\sigma$ given by:
\begin{eqnarray}
\sigma &\equiv& \frac{n_{\gamma}}{\sqrt{n_{\rm bkg}}}\\ &=&
\frac{\sqrt{T_\delta} \epsilon_{\Delta \Omega}}{\sqrt{ \Delta \Omega}}
\frac{\int A^{\rm eff}_\gamma (E,\theta) [d\phi^{\rm DM}_\gamma/dE
d\Omega] dE d\Omega}{\sqrt{ \int \sum_{\rm bkg} A^{\rm eff}_{\rm
bkg}(E,\theta) [d\phi_{bkg}/dEd\Omega] dE d\Omega}} \nonumber
\label{sensitivity}
\end{eqnarray}
where $T_\delta$ defines the effective observation time and
$\phi_{bkg}$ is the background flux. For a ${\rm \check{C}}$erenkov
apparatus, for instance, it is defined as the time during which the
source is seen with zenith angle $\theta \leq 60^{\circ}$. The
quantity $\epsilon_{\Delta \Omega} = 0.7$ is the fraction of signal
events within the optimal solid angle $\Delta \Omega$ corresponding to
the angular resolution of the instrument.  The effective detection
areas $A^{\rm eff}$ for electromagnetic and hadronic induced showers
are defined as the detection efficiency times the geometrical
detection area.  For the case of a ${\rm \check{C}}$erenkov apparatus
we have assumed a conservative effective area $A^{\rm eff} = 4 \times
10^8 \cm^2$, while for a satellite experiment we have considered
$A^{\rm eff} = 10^4 \cm^2$. Both values have been assumed independent
from $E$ and $\theta$.  Note that while the former can be increased by
adding together more ${\rm \check{C}}$erenkov telescopes, the latter
is intrinsically limited by the size of the satellite and cannot be
much greater than the fiducial value quoted here.  Finally we have
assumed an angular resolution of $0.1^{\circ}$ for both instruments,
and a total effective pointing time of 20 days for the ${\rm
\check{C}}$erenkov telescope and 30 days for the experiment on
satellite.  An identification efficiency $\epsilon$ must be taken into
account, which is one of the most important factors which have to be
studied in order to reduce the physical background level.  A ${\rm
\check{C}}$erenkov apparatus has a typical identification efficiency
for electromagnetic induced (primary $\gamma$ or electrons) showers
$\epsilon_{\rm e.m.}\sim 99 \%$ and for hadronic showers
$\epsilon_{\rm had}\sim 99 \%$. This means that only 1 hadronic shower
out of 100 is misidentified as an electromagnetic
shower. Unfortunately, this method cannot distinguish between primary
photons and electrons, which therefore represent an irreducible
background for ground-based detectors.  As far as a satellite-borne
experiment is concerned, an identification efficiency for charged
particles of $\epsilon_{\rm charged}\sim 99.997 \%$ can be assumed,
while for photons it lowers to $\epsilon_{\rm neutral}\sim 90 \%$ due
to the backsplash of high energy photons \cite{moiseev}.

We have considered the following values for the background levels. For
the proton background we use \cite{hbck}:
\begin{equation}
\frac{d \phi^h}{d\Omega dE} = 1.49 E^{-2.74} \frac{p}{\cm^2 \sec \sr \GeV} ,
\label{hadrons}
\end{equation}
while for the electron background \cite{elbck}:
\begin{equation}
\frac{d \phi^e}{d\Omega dE} = 6.9 \times 10^{-2} E^{-3.3} \frac{e}{\cm^2 \sec \sr \GeV} 
\label{electrons}
\end{equation}
and finally for the Galactic photon emission, as extrapolated by EGRET data at 
lower energies, we employ \cite{Bergstrom:98}:
\begin{equation}
\frac{d \phi^{\rm gal-\gamma}_{\rm diffuse}}{d\Omega dE}=
 N_0(l,b) \;10^{-6}\; E_{\gamma}^{\alpha} \frac{\gamma}{\cm^2 \sec \sr \GeV},
\label{dndegal}
\end{equation}
with $\alpha$ set to $-2.7$ in all the considered energy range, in lack
of data for energies higher than tens of GeV. The normalization factor
$N_0$ depends only on the interstellar matter distribution, and is
modeled as \cite{Bergstrom:98}:
\begin{equation} 
N_0(l,b)=\frac{85.5}{\sqrt{1+\left(l/35\right)^2}\;\sqrt{1+\left(b/(1.1+|l|\,0.022)\right)^2}}
 \,+\,0.5
\end{equation}
for $|l|\, \geq 30^{\circ}$ and
\begin{equation} 
N_0(l,b) =\frac{85.5}{\sqrt{1+\left(l/35\right)^2}\;\sqrt{1+\left(b/1.8\right)^2}}\,+\,0.5
\label{n0}
\end{equation}
for $|l|\,\leq 30^{\circ}$, where the longitude $l$ and the latitude
$b$ are assumed to vary in the intervals $-180^{\circ} \leq l \leq
180^{\circ}$ and $-90^{\circ}\leq b \leq 90^{\circ}$, respectively.
Finally, for the diffuse extragalactic $\gamma$ emission, as
extrapolated from EGRET data at lower energies \cite{gbck}, we use:
\begin{equation}
\frac{d \phi^{\rm extra-\gamma}_{\rm diffuse}}{d\Omega dE} 
= 1.38 \times 10^{-6} E^{-2.1} 
\frac{\gamma}{\cm^2 \sec \sr \GeV}. 
\label{gammas}
\end{equation}
If a galactic origin of high galactic latitude $\gamma$ emission is
considered, then this last estimate should be increased by about 60\%
\cite{alvaro}.

\begin{figure}[t] 
\vspace{-20pt}
\includegraphics[height=8cm,width=8cm]{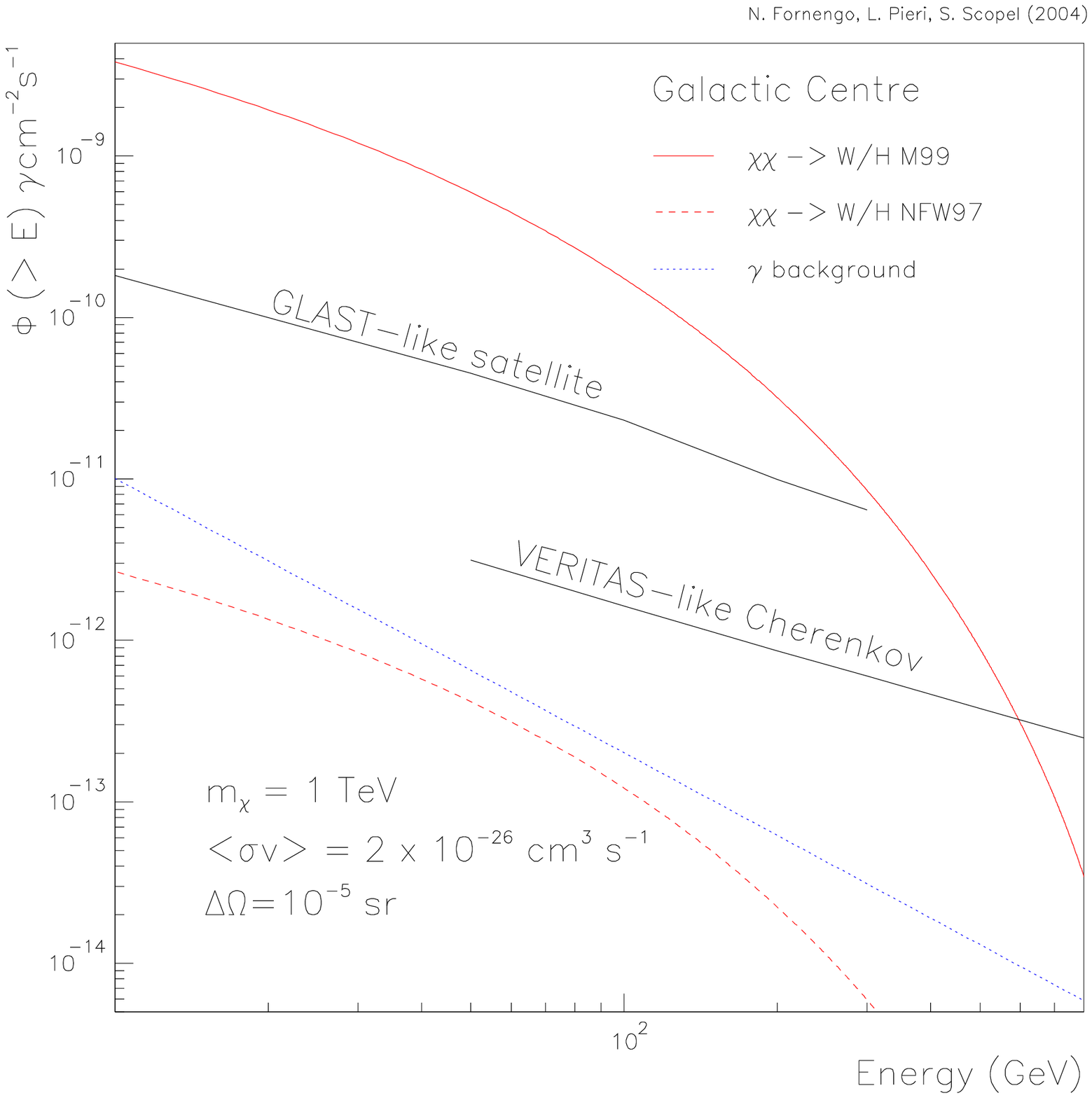}
\includegraphics[height=8cm,width=8cm]{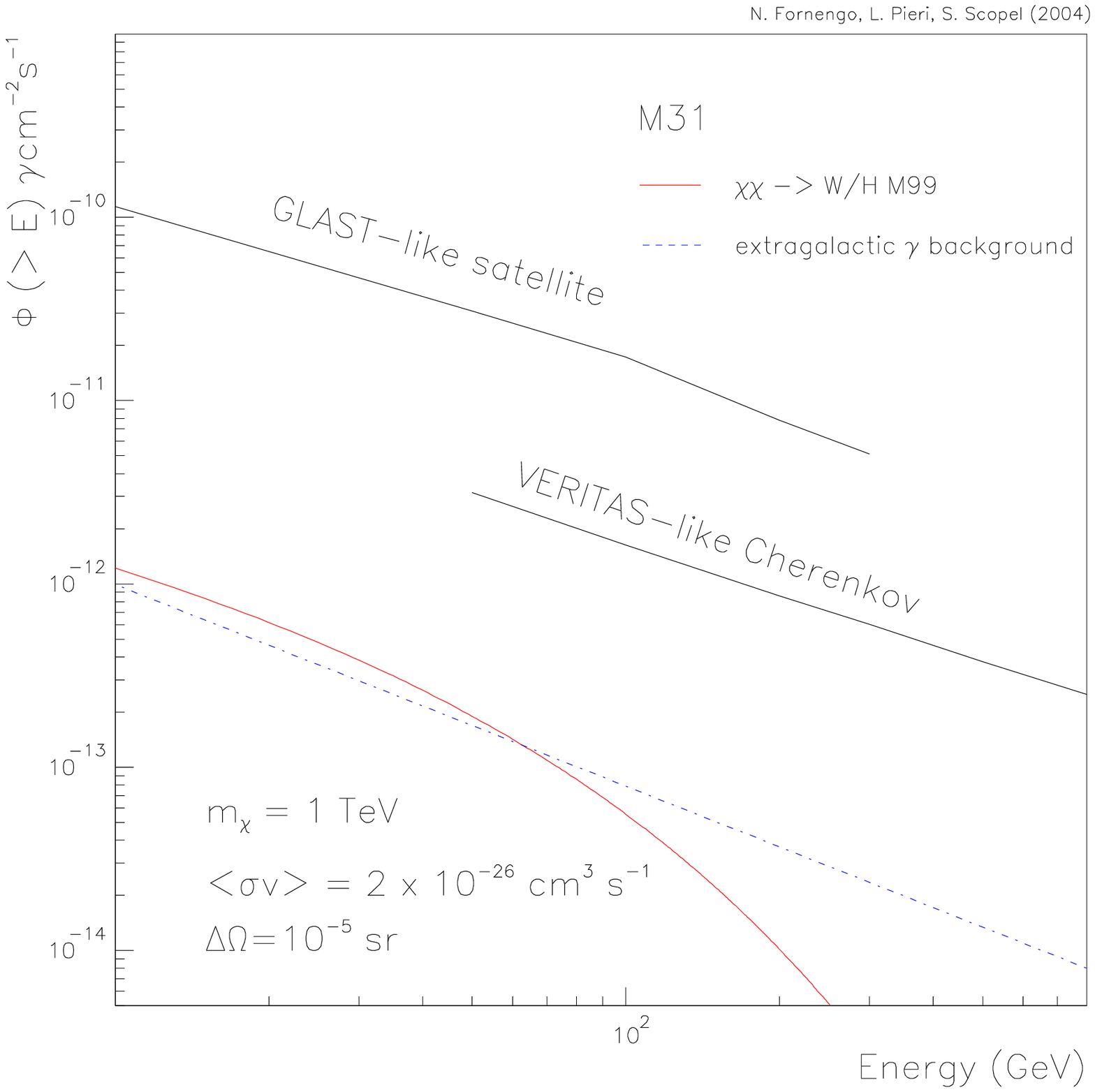}
\caption{Study of the sensitivity of an ACT detector and a satellite
borne experiment to photon fluxes from a TeV neutralino annihilation.
Solid lines denote the $5\sigma$ sensitivity curves for satellite and
${\rm \check{C}}$erenkov detectors. These curves have been calculated
according to the prescriptions given in the text. The flux expected
from the GC with a NFW97 and a M99 profile are shown in the upper
panel. The flux from M31 with a M99 profile is shown in the lower
panel.  Photon fluxes are given for $\Delta \Omega = 10^{-5}$ sr,
which is the typical detector acceptance.}
\label{fig9}
\end{figure}

Fig. \ref{fig9} shows the 5 $\sigma$ sensitivity curves for the
experimental apparata discussed above. Due to the different $\gamma$
backgrounds, the curves are slightly different in the direction of the
galactic center or toward the M31 galaxy. Also plotted for reference
is the expected integrated $\gamma$-ray flux for a SUSY model with
$m_\chi = 1 \TeV$, 50\% branching ratio of annihilation into W bosons
and 50\% into Higgs bosons (following the results of Fig. \ref{fig4a}
for the branching ratios of high mass neutralinos), and an
annihilation cross-section of $2 \times 10^{-26} \cm^{3} \sec^{-1}$
which refers to the most optimistic values of Figs. \ref{fig7} and
\ref{fig7a}. Due to our discussion in the previous Section on the
properties of $\Phi^{\rm SUSY}$, one could then consider the curve of
$\gamma$-ray flux from neutralino annihilation which we show in
Fig. \ref{fig9} as the highest spectrum of a range of curves given by
the spread of points in Figs. \ref{fig7} and \ref{fig7a}.

From Fig. \ref{fig9} and our previous discussion on the cosmological
and supersymmetric factor it therefore arises that signals from
extragalactic objects could hardly be detected. The gamma--ray
spectrum calculated for a M99 profile is two orders of magnitude
smaller than the expected sensitivities we estimate for detectors like
GLAST and about one order of magnitude smaller than the estimated
sensitivity of VERITAS. We also notice that the most optimistic
prediction for the flux we are showing in Fig.  \ref{fig9} is at the
level of the extrapolated background, a fact which by itself would
make problematic the observation of a signal from M31. Only in the
very optimistic case of a clumpy M99 matter density, the expected
signal would exceed the extrapolated background, but it would
nevertheless remain unaccessible.

In the case of a signal from the galactic center, a density profile as
cuspy as M99 (or the adiab--NFW) could be resolved by both a satellite
detector like GLAST and a ${\rm \check{C}}$erenkov telescopes with the
characteristics of VERITAS. In the case of a NFW97 profile, a potential
signal would not be accessible. Therefore, in the case of the signal
from the galactic center a density profile harder than NFW97 is required
in order to have a signal accessible to GLAST--like and VERITAS--like
detectors.

\subsection{Comparison with Recent Data}

\begin{figure}[t] 
\vspace{-20pt}
\includegraphics[height=8cm,width=8cm]{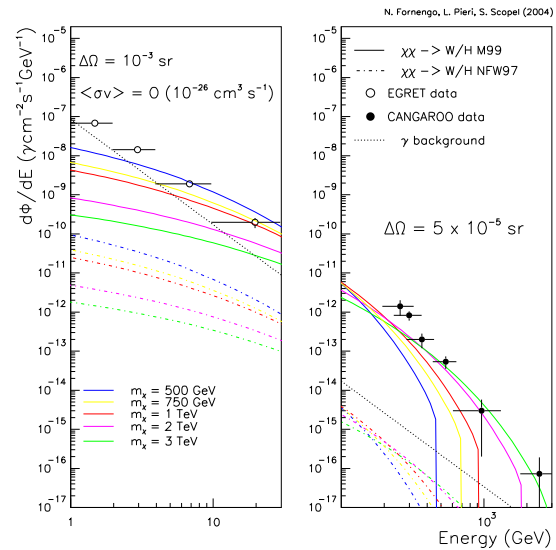}
\caption{ Differential spectrum of the photon flux expected from
neutralino annihilation in the galactic center. A 50\% branching ratio
into $W$ pairs and 50\% into $b$ quarks has been assumed. Solid lines
represent the calculation for a M99 profile for different neutralino
masses, while dashed-dotted lines show the same spectra assuming a NFW97
profile.  Dotted lines shows the extrapolated $\gamma$-ray
``conventional'' background.  Open circles (left panel) show the EGRET
results on photon flux from the galactic center, while filled circles
(right panel) show the recent data at higher energies from CANGAROO-II.
Photon fluxes are given for the corresponding typical detector
acceptance, that is for $\Delta \Omega = 10^{-3}$ sr in the left panel
and for $\Delta \Omega = 5 \cdot 10^{-5}$ sr in the right panel.}
\label{fig8}
\end{figure}

\begin{figure}[t] 
\vspace{-20pt}
\includegraphics[height=8cm,width=8cm]{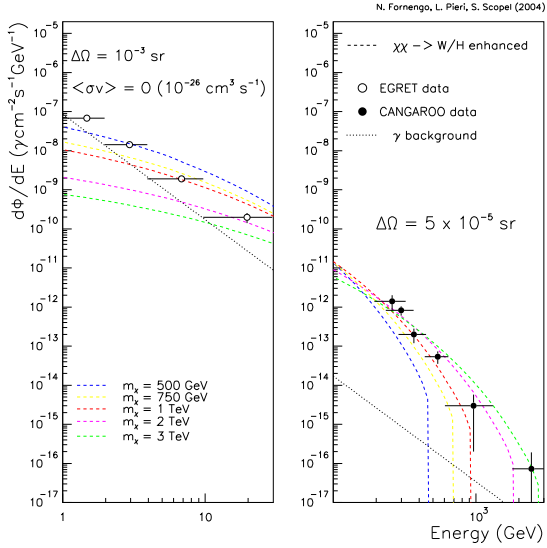}
\caption{
The same as in Fig. \ref{fig8} for a M99 profile multiplied by a factor 2.5
(dashed lines).
}
\label{fig8a}
\end{figure}

Recent experimental data taken from CANGAROO-II \cite{CANGAROO} in the
direction of the galactic center, show that the spectral shape of
photons from the GC is in excess of the extrapolated background from
standard processes. Fig. \ref{fig8} shows the CANGAROO-II data in the
right panel, and the EGRET data \cite{egret_gc} at lower energies in
the left panels. We have superimposed to the data the $\gamma$-ray
background used in our previous analysis, as well as the predicted
$\gamma$-ray spectra from high mass neutralino annihilation, for the
NFW97 and the M99 profiles. These spectra have been normalized within
a solid angle coherent with the observations. We can see that not even
a M99 profile can reproduce the observed data, as already observed in
Ref.  \cite{silk:04}.  Fig. \ref{fig8a} reproduces the same
information of Fig. \ref{fig8}, but the ``cosmological factor'' has
been enhanced by a factor 2.5 (equivalently, one could think to an
enhancement in the ``supersymmetric factor'', but this is not possible
in the effective MSSM, neither in more constrained minimal SUGRA
models which usually provide annihilation cross sections smaller than
the effective MSSM). We can see that, when appropriately boosted, the
signals from annihilation of neutralinos with mass higher than 1 TeV
have the property of matching the observed CANGAROO-II data and not being
in conflict with the EGRET data.

On the other hand, Fig. \ref{fig8} shows that it is not possible to
explain at the same time both the EGRET excess in the 1--20 GeV energy
range and the CANGAROO-II flux at energies above 250 GeV with the
spectral shape of a gamma--ray flux from neutralino annihilation. While
the EGRET spectrum can be well explained by a light neutralino in a
non--universal gaugino model \cite{lightindirect}, with $m_\chi\sim$
30--40 GeV, or by a neutralino of about 50-60 GeV \cite{morselli} in
the effective MSSM, the CANGAROO-II data require much heavier neutralinos
in order to produce photons in the hundred of GeV range: in this case,
however, the ensuing gamma--ray spectra are too low in the 1--10 GeV
range and cannot reproduce the EGRET data together with the CANGAROO-II
ones.

\begin{figure}[t] 
\vspace{-20pt}
\includegraphics[height=8cm,width=8cm]{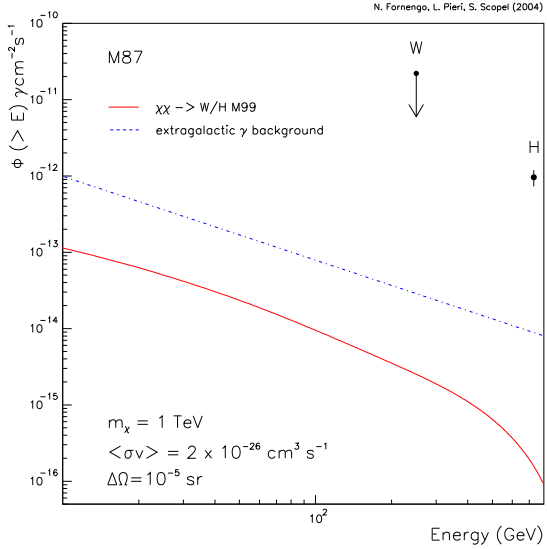}
\caption{
Integrated photon flux as expected from a TeV neutralino annihilation in
the M87 galaxy.
Photon fluxes are given for $\Delta \Omega = 10^{-5}$ sr, which is the
typical detector acceptance. 
Also shown on the figure the upper limit determined by WHIPPLE
\cite{whipplem87} and the measurement from HEGRA \cite{HEGRAm87}.
}
\label{fig10}
\end{figure}

We complete this Section by applying our method to M87 and comparing
our results with the measurements available for that galaxy, which
show a possible indication of a $\gamma$--ray excess.  This is shown
in Fig. \ref{fig10}, where one can see that our predictions are well
below the flux measured by HEGRA \cite{HEGRAm87}, even if a M99
profile is assumed. Not even a clumpy distribution, which could
enhance the predicted fluxes by at most a factor of 5, would allow us
to explain the HEGRA excess by means of neutralino annihilations in
the effective MSSM.

\section{Conclusions}

We have discussed the gamma--ray signal from dark matter annihilation
in our Galaxy and in external objects, namely the Large Magellanic
Cloud, the Andromeda Galaxy (M31) and M87. The aim of our paper was to
derive consistent predictions for the fluxes in a specific realization
of supersymmetry, the effective MSSM, and to compare the predictions
with the capabilities of new--generation satellite--borne
experiments, like GLAST, and ground-based ${\rm \check{C}}$erenkov
telescopes, for which we have used, for definiteness, the
characteristics of the VERITAS telescope.

Our results show that only the signal from neutralino annihilation at
the galactic center could be accessible to both satellite--borne
experiments and to ACTs, even though this requires very steep dark
matter density profiles toward the galactic center. A profile steeper
than NFW97 is required in order to provide signals which can reach
detectable levels. In the case of signals coming from external
galaxies, even though the extragalactic signal is larger than the
galactic contribution from neutralino annihilation, nevertheless the
absolute level of the flux is too low to allow detection with the
experimental techniques currently under development.

We have also compared our theoretical predictions with the recent
CANGAROO-II data from the galactic center and with the HEGRA data from
M87. In both cases an indication of a gamma--ray excess is present.
In the case of the CANGAROO-II data, the spectral shape is well
reproduced by a gamma--ray flux from annihilation of neutralinos
somewhat heavier than about 1 TeV, in agreement with
Ref. \cite{silk:04}. However the overall normalization of the flux
requires a boost factor of about 2.5 over the flux obtained with a
Moore et al. profile: this seems hard to obtain even in the presence
of clumps. We also showed that the agreement with the CANGAROO-II data
which is obtained with these boosted fluxes is not in contrast with
the lower--energy EGRET data from the galactic center. In addition we
showed that the spectral features of such fluxes cannot explain at the
same time both the CANGAROO-II and EGRET excess by invoking a very heavy
neutralino. Finally, we compared our predictions for the signal from
M87 with the HEGRA data and found that the predicted fluxes from
neutralino annihilation are too low to explain the HEGRA result.

\begin{figure}[t] 
\vspace{-20pt}
\includegraphics[height=8cm,width=8cm]{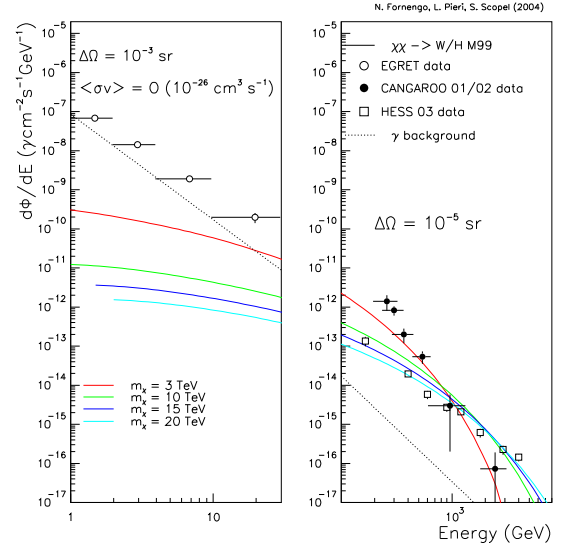}
\caption{ The same as in Fig.\ref{fig8}, including the data from HESS
\cite{hess} (see Note Added at the end of the paper). Photon fluxes are
shown for neutralino masses up to 20 TeV and for an M99 density
profile.}
\label{fig:hess}
\end{figure}

\section{Note added}
The HESS ${\rm \check{C}}$erenkov telescope \cite{hess} has recently
published new data on gamma rays from the galactic center. The
measured flux and spectrum differ substantially from previous results,
in particular those reported by the CANGAROO collaboration, exhibiting
a much harder power--law energy spectrum, with spectral index of
about $-2.2$. According to our analysis, these data, if interpreted
in terms of neutralino annihilation, would require a neutralino mass
in the range 10 TeV $\lsim m_{\chi}\lsim$ 20 TeV and an M99 profile
for the DM distribution, as shown in Fig.\ref{fig:hess}.

\acknowledgments We warmly thank E. Branchini and F. Donato for useful
discussions. We acknowledge Research Grants funded jointly by the
Italian Ministero dell'Istruzione, dell'Universit\`a e della Ricerca
(MIUR), by the University of Torino and by the Istituto Nazionale di
Fisica Nucleare (INFN) within the {\sl Astroparticle Physics Project}.

\end{document}